\documentclass[12pt,letterpaper,twoside,openright]{report} 

\usepackage[english]{babel}

\usepackage{graphicx}

\usepackage{psfrag}
\usepackage{makeidx} 
\usepackage{amsmath} 
\usepackage{bbm,amsfonts}
\usepackage{amsthm}

\theoremstyle{definition}
\newtheorem{theorem}{Theorem}[section]
\newtheorem{lemma}[theorem]{Lemma}
\newtheorem{fact}[theorem]{Fact}
\newtheorem{proposition}[theorem]{Proposition}

\newtheorem{corollary}[theorem]{Corollary}
\newtheorem{definition}[theorem]{Definition}

\newtheorem{conjecture}[theorem]{Conjecture}
\DeclareMathOperator{\tr}{Tr}

\DeclareMathOperator{\bigoh}{O}

\newcommand{\complex}{{\mathbb C}}
\newcommand{\reals}{{\mathbb R}}
\newcommand{\trace}{\tr}
\newcommand{\prob}[1]{{\rm Pr}\left[#1\right]}
\newcommand{\expct}{{\mathrm E}}
\newcommand{\size}[1]{\left|#1\right|}

\newcommand{\ket}[1]{\left|#1\right\rangle}
\newcommand{\bra}[1]{\left\langle #1\right|}
\newcommand{\braket}[2]{\langle #1 | #2\rangle}
\newcommand{\ketbra}[2]{\ket{#1}\!\!\bra{#2}}
\newcommand{\proj}[1]{\ketbra{#1}{#1}}

\newcommand{\norm}[1]{\left\|\,#1\,\right\|}
\newcommand{\trnorm}[1]{\norm{#1}_{\mathrm {tr}}}
\newcommand{\infnorm}[1]{\norm{#1}_{\infty}}
\newcommand{\fnorm}[1]{\norm{#1}_{2}}
\newcommand{\set}[1]{{\left\{#1\right\}}}
\newcommand{\st}{{\; | \;}}

\newcommand{\bias}{{\mathrm{Bias}}}
\newcommand{\stab}{{\mathrm{Stab}}}

\newcommand{\pqc}{{\bf PQC}}
\newcommand{\identity}{{\mathbb I}}

\newcommand{\ESS}{\mathcal{S}}
\newcommand{\PP}{\mathcal{P}}
\newcommand{\emm}{\mathcal{M}}
\newcommand{\HH}{\mathcal{H}}
\newcommand{\EE}{\mathcal{E}}
\newcommand{\CC}{\complex}
\newcommand{\RR}{\mathbb{R}}
\newcommand{\UU}{\mathcal{U}}
\newcommand{\ZZ}{\mathbb{Z}}
\newcommand{\LL}{\mathcal{L}}
\newcommand{\adjoint}{\dagger}

\newcommand{\ignore}[1]{}
\newcommand{\suppress}[1]{}

\newcommand{\etal}{{\it et~al.\/}}

\def\bfact{\begin{fact}}
\def\efact{\end{fact}}

\def\bv{\left( \begin{matrix}}
\def\ev{\end{matrix} \right)}

\def\be{\begin{eqnarray*}}
\def\ee{\end{eqnarray*}}
\def\bes{\begin{eqnarray}}
\def\ees{\end{eqnarray}}
\def\bess{\begin{eqnarray*}}
\def\eess{\end{eqnarray*}}
\def\nn{\nonumber}

\begin{document}

\pagestyle{empty} 

\begin{center}

\vspace*{1.0cm}
\Huge
{Approximate Private Quantum Channels}

\vspace*{1.0cm}
\normalsize
by \\
\vspace*{1.0cm}
\Large
Paul Dickinson\\
\vspace*{2.0cm}
\normalsize
A thesis \\
presented to the University of Waterloo \\ 
in fulfilment of the \\
thesis requirement for the degree of \\
Master of Mathematics\\
in \\
Combinatorics \& Optimization\\

\vspace*{2.0cm}
Waterloo, Ontario, Canada, 2006\\

\vspace*{1.0cm}
\copyright \hspace{.1ex} Paul Dickinson 2006\\

\end{center}

\cleardoublepage


\pagestyle{plain} 
\pagenumbering{roman} 
\setcounter{page}{3}

\noindent
I hereby declare that I am the sole author of this thesis.  This is a true copy of the thesis, including any required final revisions, as accepted by my examiners. 

\noindent
I understand that my thesis may be made electronically available to the public.
%

\vspace{4cm}

\noindent
Paul Dickinson
%


\newpage

\begin{center}
\Large
\textbf{Abstract}
\end{center}

This thesis includes a survey of the results known for private and approximate private quantum channels. We develop the best known upper bound for $\epsilon$-randomizing maps, $n+2\log(1/\epsilon)+c$ bits required to $\epsilon$-randomize an arbitrary $n$-qubit state by improving a scheme of Ambainis and Smith \cite{AS04} based on small bias spaces \cite{NN90, AGHP92}. We show by a probabilistic argument that in fact the great majority of random schemes using slightly more than this many bits of key are also $\epsilon$-randomizing. We provide the first known non-trivial lower bound for $\epsilon$-randomizing maps, and develop several conditions on them which we hope may be useful in proving stronger lower bounds in the future.

\newpage

\begin{center}\textbf{Acknowledgements}\end{center}

I wish to thank my supervisor, Ashwin Nayak and readers Andris Ambainis and Richard Cleve. Thanks also go out to Matthew McKague and Niel De Beaudrap for their assistance and insights, and to Lana Sheridan and Douglas Stebila for keeping me (somewhat) sane. I also wish to thank all of the other members of the IQC, especially Debbie Leung and Daniel Gottesman, for many interesting and illuminating discussions.
\newpage

\setcounter{page}{6} 

\clearpage
\thispagestyle{empty}
\cleardoublepage

\tableofcontents

%



\newpage

\pagenumbering{arabic}
 

\pagestyle{myheadings}
\markboth{Approximate Randomization of Quantum States}
\normalsize

\clearpage
\thispagestyle{empty}
\cleardoublepage

\chapter{Introduction}\label{chap:Intro}
\markright{Introduction}

\section{Preface}

Secure communication has long been a concern. The advent of computers revolutionized the field of cryptography, and quantum computation promises to do so again. Quantum computers threaten to render obsolete modern public key cryptography, while quantum cryptography offers the prospect of unconditionally secure communication. 

Traditionally, studies of communication have assumed that one would wish to transmit classical data. In a world in which quantum information is on the rise, it is natural to think about what might happen if instead one considers messages that are quantum in nature. That is, what if one should wish to transmit a quantum state in security? This thesis addresses this question. In particular, we give an exposition of the known results for private quantum channels and for approximate private quantum channels. Our contributions are to the theory of the latter. We have improved the best known upper bounds for explicit, efficient construction of approximate private quantum channels, and provide the first known non-trivial lower bound. We have also worked on establishing tighter lower bounds.

Private Quantum Channels, and their approximate variants, use a classical key to encrypt quantum states. There are several reasons for considering schemes using a classical key. While it is possible to transmit quantum states in perfect security using quantum teleportation, this requires that the communicating parties have access to the resource of shared  entanglement. Classical information is much less volatile than quantum states, being immune to decoherence, and thus is (comparatively) easily maintained. One can envision situations in which rather than having two parties communicate, one person wishes to store a state securely for future access. Perhaps quantum ``hard drive", i.e., long term storage of quantum information, will be expensive, and provided in communal storage facilities. If one does not trust the provider of such a service not to attempt to gain access to the stored information, one could store an encrypted version. In such situations it is clearly beneficial to have an easily maintained key. Further, techniques such as quantum key distribution provide classical random keys.

In a broader context, studying randomization can provide insights into the way that noise affects quantum computation, and about the connections between classical and quantum information.

\section{Notation and Conventions}

For the Pauli Eigenvectors (\ref{sec: pauli eigenvectors}), we use several notations, depending on the context. 
\be
	\ket{0} =& & \ket{+Z}\\
	\ket{1} =& & \ket{-Z} \\
	\ket{+} =& \frac{\ket{0}+\ket{1}}{\sqrt{2}} &= \ket{+X} \\
	\ket{-} =& \frac{\ket{0}-\ket{1}}{\sqrt{2}} &= \ket{-X} \\
	\ket{+i} =& \frac{\ket{0}+i\ket{1}}{\sqrt{2}} &= \ket{+Y} \\
	\ket{-i} =& \frac{\ket{0}-i\ket{1}}{\sqrt{2}} &= \ket{-Y} 
\ee

Density operators will be represented by greek characters, usually $\rho$ for mixed states, and $\phi$ and $\psi$ for pure states, though for pure states we will usually prefer $\proj{\phi}$.

Let $a$ be an $n$-bit string. Then for an operator $P$, we will say $P^0 = \identity$ and $P^1 = P$ and define
\be
	P^a &=& \bigotimes_{j=1}^n P^{a_j}.
\ee
Using the notation $\PP_n$ for Pauli operators on $n$ qubits, we note that for any Pauli operator $P \in \PP_n$, we have $P = \alpha X^a Z^b$ for some strings $a,b$ and $\alpha \in \{1,-1,i,-i\}$. We will often ignore the initial phase factor (which is reasonable, as we will most often be conjugating with $P$), and write $P = X^aZ^b$.

We use $\LL(d)$ for the space of linear operators of $d$ dimensions, and $\mathcal{U}(d)\subset \LL(d)$ for the unitary operators of dimension $d$. Thus $\PP_n \subset \mathcal{U}(2^n)$.

A state $\rho$ of dimension $d$ refers to a positive semi-definite (and hence Hermitian) operator in $\LL(d)$ with $\tr(\rho)=1$ (\cite{NC00}, p. 101).

The notation $\EE = \{\sqrt{p_i}U_i \st i=1,..., m\}$ refers to the superoperator $\EE$ that applies $U_i$ with probability $p_i$. That is, $\EE(\rho) = \sum_{i=1}^m p_i U_i \rho U_i^\dagger$.

Throughout, $n$ will refer to the number of qubits, $d$ to the dimension of a space, $m$ to the number of operators.

\section{Quantum State Randomization}

Secret communication of a quantum state can be achieved easily using teleportation if the communicating parties have the resource of shared entanglement (or are able to establish it). We consider the case where they have access only to the weaker resource of shared classical randomness, and are interested in minimizing the amount needed. 

Suppose that two parties, Alice and Bob, share a secret key $k \in \{1, ..., m\}$, and wish to securely send a quantum state $\rho$ from Alice to Bob. One strategy that they can use is to use their key to select a unitary operation $U_k$. Then Alice can encode $\rho$ by applying $U_k$ to obtain $U_k\rho U^\dagger_k$, and send it to Bob. As Bob knows $k$, he can recover the original state $\rho$ by applying $U^\dagger_k$. Suppose now that a third party, Eve, wishes to obtain $\rho$. If her knowledge of $k$ is represented as a random variable $X$ with $\prob{X=x} = p_x$, from her perspective the state being sent from Alice to Bob is
\be
	\EE(\rho)= \sum_{x=1}^N p_x U_x \rho U^\dagger_x.
\ee
In order for this to be secure, we need to require that for all states $\rho$ that Alice and Bob might wish to communicate, the resultant states $\EE(\rho)$ are indistinguishable. In general, Alice and Bob may use a more complicated encoding procedure, potentially using ancilla states or non-unitary operations. Ambainis, Mosca, Tapp and de Wolf \cite{AMTW00} introduced the following definition.

\begin{definition}\label{defn: PQC}
Let $\ESS \subseteq \HH_{2^n}$ be a set of $n$-qubit states, $l\ge n$, $\rho_a$ an $(l-n)$-qubit density matrix (ancilla state), and $\rho_0$ a $l$-qubit density matrix (target state). Let $\EE = \left\{\sqrt{p_i}U_i : 1\le i \le m\right\}$ be a superoperator; each $U_i$ is a unitary mapping on $\HH_{2^l}$, and it is applied with probability $p_i$. That is
\be
\EE(\varphi) &\equiv& \sum_{i=1}^m p_i U_i \left( \varphi \otimes \rho_a \right) U^\dagger_i.
\ee
Then $[\ESS,\EE,\rho_a,\rho_0]$ is called a {\emph Private Quantum Channel} or {\bf PQC} if and only if for each $\varphi \in \ESS$
\be
	 \EE(\varphi)& = & \rho_0
\ee
If $l=n$ (i.e. no ancilla), we call $\EE$ length preserving and we omit $\rho_a$. 
\end{definition}

Ambainis \etal, and independently Boykin and Roychowdhury \cite{BR00}, showed that $2n$ classical bits are sufficient to randomize $n$ qubits in this fashion by using a ``quantum one-time pad".  Further, it is shown by Boykin and Roychowdhury that $2n$ bits are necessary for schemes without ancillae, and the more general result including ancillae is demonstrated in \cite{AMTW00}.  Their result was generalized to more general types of channels by Nayak and Sen \cite{NS05}, and Jain \cite{J05} has considered related communication regimes involving a variety of quantum and classical communication. These results are presented in more detail in Chapter \ref{chap: PQCs}.

\section{Approximate State Randomization}

Hayden \etal \cite{HLSW04} demonstrated that by relaxing the security condition of Definition \ref{defn: PQC}, a quadratically smaller number of operators is necessary (approximately $d$ rather than $d^2$ to $\epsilon$-randomize a space of dimension $d$). The particular relaxation they introduce is the following.
\begin{definition}\label{defn: ePQC}
Let $\ESS \subseteq \HH_{2^n}$ be a set of pure $n$-qubit states, $m\ge n$, $\rho_a$ an $(m-n)$-qubit density matrix (ancilla state), and $\rho_0$ a $m$-qubit density matrix (target state). Let $\EE = \left\{\sqrt{p_i}U_i : 1\le i \le m\right\}$ be a superoperator where each $U_i$ is a unitary mapping on $\HH_{2^m}$.

$[\ESS,\EE,\rho_a,\rho_0]$ is called a $\epsilon$-Private Quantum Channel or  $\epsilon$-PQC if and only if for each $\varphi \in \ESS$
\be
	\trnorm{\EE(\varphi \otimes \rho_a) - \rho_0} &\le& \epsilon.
\ee
where $\trnorm{\cdot}$ is the trace norm (see Section \ref{sec: trnorm}).
\end{definition}

For simplicity, we will in general take $m=n$, and thus omit $\rho_a$. We note that in the case of perfect encryption, the addition of an ancilla does not decrease the required amount of randomness \cite{AMTW00, NS05}, and that none of the upper bounds known for approximate randomization make use of any ancilla. We will also assume for the remainder of this thesis that $\rho_0 = \identity/2^n$, the totally mixed state on $n$ qubits (this is in fact necessary when one considers arbitrary quantum states of a fixed dimension, see Section \ref{sec: quantum secrecy}). Throughout this work, we are interested primarily in the number of different operators one must have available, and as such, it will almost always be the case that we consider the uniform distribution over the possible keys $k$. Thus in general, we will have a set $\ESS = \{U_k : k = 1,..., |\ESS|\}$ and
\be
	\EE(\rho) &=& \frac{1}{|\ESS|}\sum_{U_k \in \ESS} U_k \rho U^\dagger_k.
\ee
Then our requirement for $\EE$ to be $\epsilon$-randomizing is that for each $\rho$,
\begin{eqnarray}\label{eqn: erand}
	\trnorm{\EE(\rho) - \frac{\identity}{2^n}} &\le& \epsilon.
\end{eqnarray}
We will use equation (\ref{eqn: erand}) as the definition for $\epsilon$-randomizing maps. 

Although the definition of $\epsilon$-randomizing maps allows for mixed states to be sent, it is important to note that security depends crucially on an adversary not possessing a purification of the state as the following shows.

Let $\Phi = \sum_{j=1}^d \frac{1}{\sqrt{d}}\ket{j}\ket{j}$ be a maximally entangled bipartite state of dimension $d$. Let $\ESS \equiv \{U_j, p_j\}_{j = 1}^m$ be a set of unitary operators of dimension $d$ with corresponding probabilities, and $\EE$ be a superoperator on the first part with $\EE(\rho) = \sum_{j=1}^m p_j U_j \rho U^\dagger_j$. Then $(\EE\otimes\identity)(\Phi)$ has rank at most $m$, and hence
\be
\trnorm{(\EE\otimes\identity) (\Phi) - \frac{\identity}{d^2}} &\le& 2(1-m/d^2)
\ee
and for $m \in o(d^2)$, this tends towards $2$ as the dimension grows large (two states are orthogonal, and thus perfectly distinguishable if they have trace distance 2).

\section{One Qubit Example}\label{sec: one qubit example}

With 2 bits of key, we can completely randomize a single qubit by applying one of the four 1-qubit Pauli operators with equal probability. The demonstration is straightforward. Any density matrix $\rho$ can be written as 
	\bes
		\rho &=& \frac{\identity+u\cdot (X, Y, Z)}{2} \label{eqn: dm}
	\ees
for some $u\in \RR^3$, $|u| \le 1$ (\cite{NC00}, p.105). Further, using the commutation relations for Pauli operators (Section \ref{sec: pauli group}), for any $P \in \{\identity, X,Y,Z\}$, 
	\be
		\frac{1}{4}\left(\identity P \identity + XPX + YPY + ZPZ\right) &=& \begin{cases}
			\identity & \text{if $P=\identity$}\\
			0 & \text{otherwise.}
			\end{cases}
	\ee
Thus by linearity and taking $u = (u_1,u_2,u_3)$,
	\be
		\lefteqn{\frac{1}{4}\left( \sum_{P\in\PP_1} P\left(\frac{\identity+u\cdot (X, Y, Z)}{2}\right) P \right)}\\
		&=& \frac{1}{2}\sum_{P\in\PP_1}  P \identity P + \frac{u_1}{2}\sum_{P\in\PP_1}  P X P
		+  \frac{u_2}{2}\sum_{P\in\PP_1}  P Y P +  \frac{u_3}{2}\sum_{P\in\PP_1}  P Z P\\
		&=&\frac{\identity}{2}. 
	\ee

What if instead of allowing four unitary operators, we allow a uniform selection among only three unitary operators, $U_1, U_2, U_3$?

\begin{figure}[htbp]
	\centerline{	
	\includegraphics[scale=0.67]{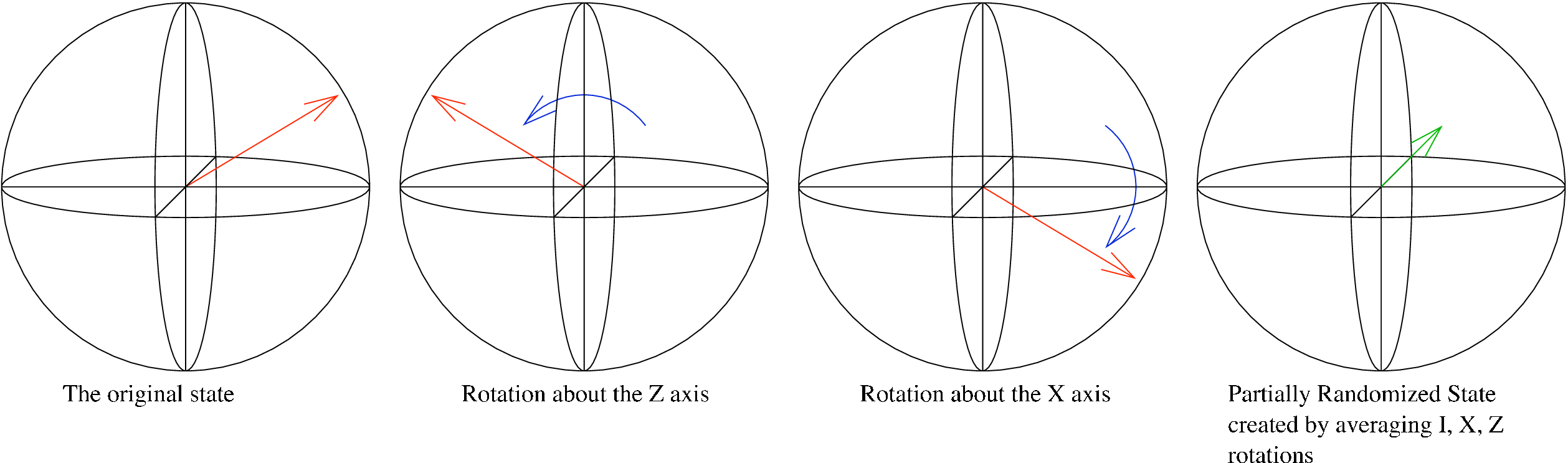}
	}
	\caption{Randomizing a single qubit using $\{I,X,Z\}$. After the randomizing operation, the state vector is contracted toward the totally mixed state.}
	\label{fig: 1/3 randomization}
\end{figure}

By the unitary invariance of the trace norm (Lemma \ref{lemma: unitarily invariant}, p.\pageref{lemma: unitarily invariant}), we can multiply each of the $U_i$ by $U_1^\dagger$ and thus determine that without loss of generality, we can take $U_1 = \identity$. As quantum states are equivalent up to global phase, we then consider the effect of applying the randomizing operator $\EE = \left\{\sqrt{\frac{1}{3}}\identity, \sqrt{\frac{1}{3}}U_2, \sqrt{\frac{1}{3}}U_3\right\}$ to $\ket{\psi}$, an eigenvector of $U_2$. Clearly both $\identity$ and $U_2$ will have no effect on $\ket{\psi}$, (and for this reason, there are no interesting schemes with only two unitaries) so we will be left with
\be
	\EE(\proj{\psi}) &=& \frac{2}{3}\proj{\psi} + \frac{1}{3}U_3\proj{\psi}U_3^\dagger.
\ee
Writing $\rho = \EE(\proj{\psi})$, by the triangle inequality, we have 
\bes
\trnorm{\proj{\psi} - \frac{\identity}{2}} &\ge& \trnorm{\proj{\psi}-\rho} + \trnorm{\rho - \frac{\identity}{2}}\label{eqn: tri}
\ees
and the left hand side of (\ref{eqn: tri}) is equal to 1.
\bes
	\trnorm{\proj{\psi}-\rho}  &=& \trnorm{\frac{1}{3}\proj{\psi} - \frac{1}{3}U_3\proj{\psi}U_3^\dagger}\nonumber\\
	&\le& \frac{2}{3} \label{eqn: tri2}.
\ees
Putting (\ref{eqn: tri}) and (\ref{eqn: tri2}) together, we have
\be
	\trnorm{\rho - \frac{\identity}{2}} &\ge& \frac{1}{3}.
\ee
Thus we see that for 3 operators, we cannot do better than $\frac{1}{3}$-randomizing of a qubit.  We can achieve the $\frac{1}{3}$ bound simply by taking any three Pauli operators, for example $\identity, X, Z$. Figure \ref{fig: 1/3 randomization} gives a geometric representation on the Bloch sphere of this scheme.

To prove that it is $\left(\frac{1}{3}\right)$-randomizing, take an arbitrary density matrix as in (\ref{eqn: dm}) and apply $\EE$.
\be
	\EE(\rho) &=& \frac{1}{6}\left(\identity + u_1 X + u_2 Y + u_2 Z\right) + \frac{1}{6}\left(\identity + u_1 X - u_2 Y - u_2 Z\right) + \frac{1}{6} \left(\identity - u_1 X - u_2 Y + u_2 Z\right)\\
	&=& \frac{\identity}{2} + \frac{u_1}{6}X - \frac{u_2}{6}Y + \frac{u_3}{6}Z 
\ee
then we have 
\bes
	\trnorm{\EE(\rho) - \frac{\identity}{2}} &=& \frac{1}{6}\trnorm{u_1X - u_2Y + u_3Z} \label{eqn: tn}
\ees
As $\left(u_1X - u_2Y + u_3Z\right)$ has eigenvalues $\pm |u|$ and $|u|\le 1$, (\ref{eqn: tn}) gives us
\be
	\trnorm{\EE(\rho) - \frac{\identity}{2}} &\le& \frac{1}{3}.
\ee
Thus the randomization scheme consisting of applying one of $\{\identity, X, Z\}$ with equal probability is optimal for 1-qubit schemes using exactly 3 operators.

Geometrically, we can easily visualize the one-qubit case. The image $\EE(\rho)$ is obtained by adding the vectors $p_i U_i \rho U_i^\dagger$. Then the trace norm  $\trnorm{\EE(\rho)-\frac{\identity}{2}}$ is the maximum over states $\rho$ of the length of the vector $\EE(\rho)$ in the Bloch sphere.

\section{Results}

We have shown two upper bounds for the amount of randomness required for  $\epsilon$-randomizing maps in the trace norm.

In Chapter \ref{chap: probabilistic} we show by a probabilistic argument similar to that of Hayden \etal~\cite{HLSW04} that a sufficiently large random set of Pauli operators is $\epsilon$-randomizing with high probability.
\begin{theorem}\label{thm: trnorm}
For every $d$, and $\epsilon>0$, a sequence of $\frac{37d}{\epsilon^2}\log\left(\frac{15}{\epsilon}\right)$ $d$-dimensional unitary operators selected at random according to the Haar measure is $\epsilon$-randomizing with probability at least $1-e^{-d/2}$. 
Thus $\log(d) + \bigoh(\log(1/\epsilon))$ bits of key suffice to randomize arbitrary $d$ dimensional states. 
Furthermore, for $d = 2^n$, the unitaries may be selected uniformly at random from $\PP(n)$.
\end{theorem}

In Chapter \ref{chap: explicit} we provide the best known explicit construction (Theorem \ref{thm: construction}) based on the work of Ambainis and Smith \cite{AS04}. 
\begin{theorem}\label{thm: construction}
	There exists an efficiently constructible $\epsilon$-randomizing map on $n$ qubits, and requiring $2^n/\epsilon^2$ operators, or alternatively, $n+\log(1/\epsilon^2)$ bits of key.
\end{theorem}

Lower bounds, Chapter \ref{chap: lower bounds}, have proven more difficult to establish. We have shown the following weak lower bound on the class of length-preserving $\epsilon$-randomizing channels that use only Pauli operators.
\begin{theorem}\label{thm: weak bound}
	Let $\ESS \subset \{0,1\}^n \times \{0,1\}^n$ be a set of binary strings $(a,b)$, such that the map
	\be
		\EE(\rho) &=& \frac{1}{|\ESS|} \sum_{(a,b)\in \ESS} X^aZ^b \rho Z^bX^a
	\ee
	is $\epsilon$-randomizing. Then $\ESS$ is an $\epsilon$-biased space, and thus
	\be
	|\ESS| &\ge& \Omega\left(\frac{n}{\epsilon^2 \log \left(\frac{1}{\epsilon}\right)} \right).
	\ee
\end{theorem}
Our bound introduces a dependence on $\epsilon$, and is better than the naive bound for $2^{-n} \le \epsilon \le 2^{-n/2}$, where the bound becomes $\Omega\left(\frac{1}{\epsilon^2}\right)$.

We have also done work on establishing a tighter lower bound by finding additional conditions that $\epsilon$-randomizing maps must satisfy. In particular, we show the following theorem specifying the action of $\epsilon$-randomizing maps on stabilizer states.
\begin{theorem}\label{thm: stab condition}
	For every stabilizer group $\hat{G}= \langle g_1,...,g_n\rangle$, let $G$ be the matrix with rows $g_1,..., g_n$. Let $\ESS \subset \{0,1\}^n \times \{0,1\}^n$ be a set of binary strings $(a,b)$ such that
	\be
		\EE(\rho) &=& \frac{1}{|\ESS|} \sum_{(a,b)\in \ESS} X^aZ^b \rho Z^bX^a
	\ee
	is $\epsilon$-randomizing. Let $G(\ESS)$ be the distribution on $\{0,1\}^n$ with $p(k) = |G^{-1}(k)|/|\ESS|$, where $G^{-1}(k)$ is the pre-image of $k$.  Then $G(\ESS)$ must be at most $\epsilon$-away from uniform.
\end{theorem}

In other words, in order for a set $\ESS$ to be $\epsilon$-randomizing, for every stabilizer state, the image of the linear transformation given by its generator matrix $G$ applied to $\ESS$ must be nearly uniform. This condition is the strongest one that we have discovered, and implies the weak lower bound of Theorem \ref{thm: weak bound}.  

\section{Related Work}
A map is said to be $\epsilon$-randomizing in the $2$-norm if for every state $\rho$, 
\be
\fnorm{\EE(\rho) - \frac{\identity}{d}} &\le&  \frac{\epsilon}{\sqrt{d}}.
\ee
Kerenidis and Nagaj \cite{KN05} studied maps that act independently on each qubit, and found the minimal amount of entropy required to create  independently acting $\epsilon$-randomizing maps in the $2$-norm for each positive $\epsilon$. They determine the optimal approximate encryption schemes for 1 qubit with any given entropy and also show that the constructions of $\epsilon$-randomizing maps based on small-biased spaces of \cite{AS04} do not work in the $\infty$-norm.

Christoph Dankert's thesis \cite{D06} on approximate 2-designs and mutually unbiased bases is related to this work. In his terminology, an $\epsilon$-randomizing map is an approximate 1-design for quantum states.  This view may give some intuition into how to improve lower bounds.

Dealing as it does with quantum state randomization, there are obvious parallels to be drawn with cryptography, both classical and quantum, and also on the precise nature of the relationship between quantum and classical information.

\clearpage
\thispagestyle{empty}
\cleardoublepage

\chapter{Private Quantum Channels}\label{chap: PQCs}
\markright{Private Quantum Channels}

\section{Quantum Secrecy}\label{sec: quantum secrecy}

Suppose we have two parties, Alice and Bob, sharing some key $k$ in a set $\{1, ..., m\}$ with respective probabilities $p_k$ for the $i$'th key,  and Alice wishes to send some state $\rho$ from a set $\ESS$ to Bob securely. A general encryption procedure would be for Alice to append to $\rho$ some ancilla qubits in some state $\rho_{a}$, which may depend upon the key $k$, and then perform some unitary operation $U_k$ depending on the key $k$ on the combined state $\rho\otimes\rho_a$ \cite{NS05}. She would then send the state she has created to Bob, who, knowing $k$ would perform $U_k^\dagger$ and remove the ancilla to recover the state $\rho$. 

From the perspective of Eve who does not know $k$, the state being sent will be a probability distribution over the possible ciphertexts:
\be
	\EE(\rho\otimes\rho_a) &=& \sum_{i=1}^m p_i U_i (\rho \otimes \rho_a) U_i^\dagger.
\ee
In order for this procedure to be information-theoretically secure, an eavesdropper who obtains the encrypted state but knows only the distribution from which $k$ is taken should not be able to gain any information about the state $\rho$. In other words,  for any two states $\rho_1, \rho_2 \in \ESS$, the encrypted states $\EE(\rho_1\otimes\rho_a)$ and $\EE(\rho_2\otimes\rho_a)$ must be indistinguishable by any process; their density matrix representations must be equal (\cite{NC00}, Section 9.2). We will write $\EE(\rho) = \rho_0$. 



Recall the definition of {\pqc}s, Definition \ref{defn: PQC} (p.\pageref{defn: PQC}). Observe that if a scheme works for some set of states $\ESS$, by linearity it will also work for convex combinations of those states, and hence if we have a \pqc~on pure states, it will also work for mixed states. As we will generally be interested in the case $\ESS = \HH_{2^n}$, this means that the totally mixed state $\frac{\identity}{2^n}$ is a valid message, and it is clear that  for any length preserving $\EE$, we must have $\rho_0 = \frac{\identity}{2^n}$.

Also note that although Eve cannot gain any information about the state that was sent via a \pqc, she can still destroy it by measuring (or simply not transmitting it). This is equivalent to classically jamming a channel, and cannot be avoided (though it can be detected \cite{BHLMO04}).

\section{Classical One-Time Pad}
The classical one-time pad or Vernam cipher is well-known to provide perfect encryption for classical messages. To encrypt a message $M$, one first compresses it to its entropy $n$,  and then uses a key $k$ of the same length. The cipher text is obtained by taking the exclusive-or of the compressed message $M'$ and $k$. If $k$ is uniformly distributed, so is $M'\oplus k$. To recover $M'$, one takes the exclusive-or of the cipher text and the key to obtain 
$(M'\oplus k) \oplus k = M' \oplus (k \oplus k) = M'$, and then reverse the compression procedure. It is also known \cite{S49} that this result is optimal for classical communication. In order to obtain a uniformly distributed cipher text, one must use at least $n$ random bits (skipping the compression step simply means that more key will be used). It is noteworthy that this result does not translate directly to the quantum world; with a quantum channel and 2-way classical communication, one can perform a quantum key distribution (or more accurately, key-growing) protocol such as BB84 \cite{BB84} using much less shared key, and then use the output of such a protocol to encrypt the original message. 

\section{Quantum One-Time Pad}\label{sec: qotp}

One can generalize the classical one-time pad to the quantum world in the following manner (\cite{AMTW00}, \cite{BR00}). Instead of randomly flipping the bit value as in the classical case, one applies a random Pauli operator to each qubit. We have shown the one qubit case in Section \ref{sec: one qubit example}. Generalizing to multiple qubits is straightforward. 
We make use of the following technical lemma (\cite{AMTW00}, Lemma 4.4).
\begin{lemma}[\cite{AMTW00}, Lemma 4.4]\label{lemma: off-diagonal terms}
	 Suppose that $\EE(\proj{\phi}\otimes\rho_a) = \rho_0$ whenever $\ket{\phi}$ is a tensor product of $n$ qubits. Then $\EE(\ketbra{x}{y}\otimes \rho_a) = 0$ whenever $x$ and $y$ are different $n$-bit strings.
\end{lemma}

\begin{theorem}\label{thm:egpqc}
	$[\HH_{2^n}, \PP_n, \identity/{2^n}]$ is a \pqc.
\end{theorem}
\begin{proof}
	The proof is by induction. We first observe that the claim is true for $n=1$ (as shown in Section \ref{sec: one qubit example}). We refer to the one-qubit scheme as $\EE'$. Then we claim that if $\EE$ is a \pqc~for $\HH_{2^{k-1}}$, then $\EE\otimes\EE'$ is a \pqc~for $\HH_{2^k}$.
	Consider an arbitrary pure state 
\be 
	 \ket{\phi} = \sum_{x\in\{0,1\}^k, y\in\{0,1\}} \alpha_{xy}\ket{x}\otimes\ket{y}.
\ee
Applying $\EE\otimes\EE'$ to $\ket{\phi}$ yields
	\bes
		(\EE\otimes\EE')(\proj{\phi}) 
		&=& (\EE\otimes\EE')\left(\sum_{x,x'\in\{0,1\}^k, y,y'\in\{0,1\}} \alpha_{xy}\alpha_{x'y'}^*
			\ketbra{x}{x'}\otimes\ketbra{y}{y'}\right) \nonumber\\
		&=& \sum_{x,x', y,y'}\alpha_{xy}\alpha_{x'y'}^*
			\EE(\ketbra{x}{x'})\otimes\EE'(\ketbra{y}{y'}) \nonumber\\
		&=& \ \sum_{x,x', y,y'}\alpha_{xy}\alpha_{xy}^*
			\EE(\ketbra{x}{x}) \otimes \EE(\ketbra{y}{y})  \label{eqn: offdiag}\\
		&=& \sum_{x, y}|\alpha_{xy}|^2 
			\frac{\identity_{2^{k-1}}}{2^{k-1}} \otimes \frac{\identity_2}{2}\nonumber\\
		&=& \frac{\identity_{2^k}}{2^k}\nonumber
	\ees
where (\ref{eqn: offdiag}) follows by Lemma \ref{lemma: off-diagonal terms}. By convexity, $\EE\otimes\EE'$ is also a randomizing map for any mixed states.
\end{proof}
Ambainis \etal~\cite{AMTW00} prove a slightly more general version that also demonstrates that in fact the concatenation of arbitrary {\pqc}s is a \pqc. As choosing a Pauli operator takes 2 bits, this scheme requires $2n$ bits of private key. 
%


It turns out that if we restrict to sending only qubits with only real amplitudes, for $n=1,2,3$, $n$ bits of randomness are sufficient to totally randomize any state, but this is equivalent to giving an adversary some information about each qubit. Similarly, there are {\pqc}s for which $\rho_0$ is not proportional to $\identity$, but these correspond to rather specialized sets of possible messages. For further details, see \cite{AMTW00}.

\section{Lower Bound on Entropy of {\pqc}s}
We have demonstrated that $2n$ bits of entropy is sufficient to securely send any $n$-qubit state, and in this section we show that in fact this is the minimum amount of entropy needed to encrypt arbitrary $n$-qubit states. This result was first shown by Boykin and Roychowdhury in \cite{BR00} for {\pqc}s without ancillae. The more general case allowing a fixed ancilla was shown in \cite{AMTW00}  by Ambainis, Mosca, Tapp and de Wolf, and extended to a key-dependent ancilla by Nayak and Sen \cite{NS05}. Jain \cite{J05} has shown several related bounds.

Boykin and Roychowdhury \cite{BR00} showed that a superoperator $\EE = \{\sqrt{p_k}U_k\}$ that totally randomizes arbitrary $n$-qubit messages without the use of an ancilla must have the property that $\{U_k\}$  spans $\LL(2^n)$, and since any spanning set  for $\LL(2^n)$ must have dimension at least $2^{2n}$ over $\RR$, $2n$ is a lower bound on the number of necessary bits of key. The proof here is a simplified version due to Ambainis Mosca, Tapp and de Wolf \cite{AMTW00}, and uses the following theorem  (\cite{NC00}, Theorem 8.2). 

\begin{lemma}\label{lemma: unitary relation}
	Suppose two operators $\EE = \{\sqrt{p_i}U_i\}_{i=1}^m$ and $\EE' = \{\sqrt{q_j}U'_j\}_{j=1}^m$ (one of them possibly padded with $0$ operators) are such that for every state $\rho$, $\EE(\rho) = \EE'(\rho)$. Then there is some unitary $A\in \LL(m)$ such that 
	\be
		\sqrt{p_i}U_i &=& \sum_{j=1}^m A_{i,j}\sqrt{q_j}U'_j.
	\ee
\end{lemma}


\begin{theorem} If $[\HH_{2^n}, \EE = \{\sqrt{p_i}U_i : 1\le i \le m\}, \identity_{2^n}]$ is a \pqc\ then the Shannon entropy $H(p_1,...,p_N) \ge 2n$.
\end{theorem}
\begin{proof}
Let $\EE' = \left\{\frac{1}{\sqrt{2^{2n}}}\overline{\sigma_x} : x \in \{0,1,2,3\}^n\right\}$ be the superoperator defined in Theorem \ref{thm:egpqc}. let $k=\max\{2^{2n},m\}$. Since  $\EE(\rho)=\EE'(\rho)=\identity_{2^n}$ for all states $\rho$,  $\EE$ and $\EE'$ are unitarily related as in Lemma \ref{lemma: unitary relation}. So, there is a $k\times k$ unitary  matrix $A$ such that $\forall 1 \le i \le m$ we have 
\begin{equation}
\sqrt{p_i}U_i = \sum_{x\in\{0,1,2,3\}^n} A_{ix} \frac{1}{\sqrt{2^{2n}}}\overline{\sigma_x}.
\end{equation}
The set of $2^n\times 2^n$ matrices forms a $2^{2n}$ dimensional vector space with inner product $\langle M , M' \rangle = \tr(M^\dagger M' )/2^n$, and induced norm $\norm{M} = \sqrt{\langle M, M \rangle}$ (This is simply a re-normalization of the $2$-norm, Section \ref{sec: 2-norm}, p.\pageref{sec: 2-norm}). Note that $\norm{U}=1$ for unitary $U$ and the set of all $\overline{\sigma_x}$ forms an orthonormal basis for this space (Lemma \ref{lemma: paulis orthogonal basis}, p.\pageref{lemma: paulis orthogonal basis}), so we have:
\begin{equation}
	p_i = \left\|\sqrt{p_i}U_i\right\|^2 =  \left\|\sum_x A_{ix}\frac{1}{\sqrt{2^{2n}}}\overline{\sigma_x}\right\|^2 = 
	\frac{1}{2^{2n}}\sum_x \left|A_{ix}\right|^2 \le 2^{-2n}.
\end{equation}
Thus we have $m\ge 2^{2n}$ and $H(p_1, ... , p_m) \ge 2n$.
\end{proof}

In general, one might imagine that the use of an extra ancilla state (which Bob does not need to know the state of) in addition to classical randomness might reduce the amount of entropy needed, but as was shown in \cite{AMTW00}, this is not the case. Nayak and Sen showed that the even in the  most general setting in which the ancilla may depend on the key, the same lower bound holds \cite{NS05}.

The method used to prove this in \cite{AMTW00, NS05} is to first show that any \pqc~on $n$ qubits can be transformed into a \pqc~on the set of classical states of length $2n$ (i.e., $\{\ket{j} \st j = 1,..., 2^{2n}\}$) and then prove a ``quantum analog" of Shannon's Noiseless Coding Theorem \cite{S49}, i.e., demonstrate that in order to privately send $2n$ classical bits via a \pqc, one must use $2n$ bits of entropy.

Transforming a \pqc~$\EE$ on $n$ qubits into a \pqc~$\EE'$ on $2n$ bits is achieved by encoding pairs of bits as Bell states, and then randomizing one qubit of each using $\EE$.  As randomizing half of a Bell pair leaves the whole pair in the totally mixed state this gives us a way to securely transmit $2n$ classical bits.

\begin{theorem}
Let $\mathcal{C}_{2n} = \{\ket{j} \st j=1, ..., 2^{2n}\}$ be the set of classical messages on $2n$ bits. If $[\HH_{2^n}, \EE, \rho_a, \identity_{2^n}]$ is a \pqc~with $\EE = \{\sqrt{p_i}U_i \st i=1,...,m\}$, there exists $\EE' = \{\sqrt{p_i}U'_i \st i=1,...,m\}$ such that $[\mathcal{C}_{2n},\EE', \rho_a, \identity_{2^{2n}}]$ is a \pqc.
\end{theorem}
We give a sketch of the proof, and refer the reader to \cite{AMTW00} for details.
	For simplicity of notation, assume no ancilla is used. Its addition does not affect the proof in any way. Divide $\ket{j}$ into $n$ pairs of qubits, $(i,n+i)$, $i=1,..., n$. Map each pair to a singlet state as follows
	\be
		\ket{00} &\mapsto& \frac{\ket{00}+\ket{11}}{\sqrt{2}}\\
		\ket{01} &\mapsto& \frac{\ket{00}-\ket{11}}{\sqrt{2}}\\
		\ket{10} &\mapsto& \frac{\ket{01}+\ket{10}}{\sqrt{2}}\\
		\ket{11} &\mapsto& \frac{\ket{01}-\ket{10}}{\sqrt{2}}\\
	\ee
	Note that applying a Pauli matrix to the first (or second) qubit of each of these states takes it to one of the others, and that applying a random Pauli chosen uniformly to the second qubit of any pair takes it to the totally mixed state $\frac{\identity_4}{4}$.
	Define $U'_i = \identity_{2^n} \otimes U_i$. Then $\EE'$ maps each pair of qubits $(i, n+i)$ to the totally mixed state (via a proof similar to that of , and hence $\EE'(\proj{j}) = \frac{\identity}{2^{2n}}$ for each $\ket{j}\in \mathcal{C}_{2n}$.

As the states of $\mathcal{C}_{2n}$ correspond to the possible $2n$-bit strings, this shows that any \pqc~for arbitrary $n$-qubit states can be used to perfectly encrypt arbitrary $2n$-bit classical messages. Shannon's Noiseless Coding Theorem  \cite{S49} implies that in order to securely classically communicate $n$ bits of classical information, one must use a key of at least $n$ bits of entropy. While the theorem does not carry directly into the world of quantum communication (and is violated in the case of 2-way communication), one can prove the following analogous statement \cite{AMTW00, NS05}.

 \begin{theorem}
 	Suppose $\rho$ is an $n$-qubit state, $U_k$ is a unitary operator on $d \ge 2^n$ dimensions, and $\rho_k$ is a density operator of $d/2^n$ dimensions (a key-dependent ancilla state).  Let $\EE = \left\{\left(\sqrt{p_k}U_k, \rho_k\right)\right\}_{k=1}^m$ be the superoperator that applies $U_k$ to $\rho\otimes \rho_k$ with probability $p_k$. If $[\HH_{2^n}, \EE, \rho_0]$ is a \pqc, then
	\be
		H(p_1, ..., p_m) &\ge& 2n
	\ee
where $H(p_1,..., p_m)$ is the Shannon entropy of the distribution of the $p_k$'s.
 \end{theorem}

\begin{corollary}
	The quantum one-time pad of Theorem \ref{thm:egpqc} is optimal. That is, any other \pqc~on $n$ qubits uses at least $2n$ bits of entropy.
\end{corollary}

We have a complete characterization of {\pqc}s for $n$-qubit systems. In fact the same results generalize to $d$-dimensional systems by using the Heisenberg-Weyl operators (Section \ref{sec: Heisenberg Weyl})~rather than Paulis.

\clearpage
\thispagestyle{empty}
\cleardoublepage

\chapter{Probabilistic Constructions of Approximate Randomizing Maps}\label{chap: probabilistic}
\markright{Probabilistic Constructions of Approximate Randomizing Maps}

\section{Why Probabilistic Bounds?}
Given that we have explicit constructions that match the best bounds we have obtained using probabilistic bounds, one might ask what the interest of these bounds is, especially for the case of random matrices selected according to the Haar measure, which are not efficiently constructible (they generally require an exponentially large amount of information to specify). They are interesting for two main reasons. The first is that they not only show the existence of randomization schemes, but also, by the addition of only a small amount of extra randomness can show that the overwhelming majority of sets of random operators are randomizing (or in the case of the trace norm, most sets of Pauli operators).  The second point is that by developing these techniques, we exhibit tools that may be useful in other analyses.

\section{Infinity Norm}
Hayden \etal~\cite{HLSW04} showed that by relaxing the security requirement on Private Quantum Channels by allowing a small amount of information to leak, that is, allowing messages to be encrypted to states that are ``very close" rather than identical, one can reduce the amount of private random key needed to encrypt a quantum state by a factor of two.

Recall the definition of $\epsilon$-PQCs noted in (\ref{eqn: erand}) on page \pageref{eqn: erand}. Observe that this definition is in the trace norm. Hayden \etal~gave results in the infinity norm. In particular, they showed the following theorem.

\begin{theorem}\label{thm: infnorm}
For every $d$, and $\epsilon>0$, there is a set $S$ of $m=134d\log d/\epsilon^2$ unitary operators of dimension $d$ such that for every pure state $\ketbra{\varphi}{\varphi}$
\bes
\infnorm{\sum_{U \in S} U\ketbra{\varphi}{\varphi}U^\dagger - \frac{\identity}{d}} &\le& \frac{\epsilon}{d}.
\ees
\end{theorem}

We observe that the maps described in this way are in fact $\epsilon$-randomizing in the following simple lemma.

\begin{lemma}
Any completely positive trace-preserving (CPTP) map $\EE: \mathcal{B}(\CC^d) \rightarrow \mathcal{B}(\CC^d)$ satisfying $\infnorm{\EE(\ketbra{\varphi}{\varphi}) - \frac{\identity}{d}} \le \epsilon/d$ is $\epsilon$-randomizing.
\end{lemma}
\begin{proof}
This follows from the definitions of trace and infinity norms. On normal matrices, the former corresponds to the sum of the magnitudes of the eigenvalues, and the latter to the largest eigenvalue. Clearly if the largest eigenvalue is bounded by $\epsilon/d$, then the sum of the magnitudes of the eigenvalues is bounded by $\epsilon$. For further discussion of norms, refer to Appendix \ref{app: norms}.
\end{proof}

The proof of Theorem \ref{thm: infnorm} proceeds in several steps. The first step is to prove the following lemma of Hayden \etal.
\begin{lemma}[\cite{HLSW04}]\label{lemma: cramer}
	Let $\varphi$ be a pure state, $P$ a rank $p$ orthogonal projector and let $(U_j)_{j= 1}^m$ be a sequence of dimension $d$ unitary operators selected independently and uniformly according to the Haar measure. Then there exists a constant $C \ge (6\ln2)^{-1}$ such that for $0 < \epsilon < 1$
	\be
	\prob {\left| \frac{1}{m}\sum_{j=1}^m \tr \left( U_j \varphi U_j^\dagger P\right) - \frac{p}{d}\right| \ge \frac {\epsilon p}{d} } &\le& 2 \exp (-Cmp\epsilon^2).
	\ee
\end{lemma}
This bounds the probability that for a fixed state $\varphi$, a sequence of $m$ random unitary operators fails to randomize $\varphi$. This is done using a tail inequality known as Cram{\`e}r's Theorem (see for example \cite{DZ93}). As this lemma is not used in our argument, we omit the proof, and refer the interested reader to \cite{HLSW04}.

Next, the approximate randomizing property is extended to all states by showing that it suffices to randomize a suitable set of finitely many pure states through a net construction (Section \ref{sub: enet}). This allows us to work with only a finite number of states. 

Finally, we apply a union bound to show that for such a random sequence of unitaries, the probability that every state in the finite set is randomized is non-zero. This then implies that there is a set of $m$ unitaries that approximately randomizes every state.

\subsection{Discretization of State Space}\label{sub: enet}

In order to show that we need only consider a finite number of states, we need to show that if two states $\ketbra{\phi}{\phi}$ and $\ketbra{\psi}{\psi}$ are ``close enough" together, then randomizing one implies that the other is also randomized.  We must first show that there is in fact a finite set of states $\mathcal{M}$ such that every state is close to some element of $\mathcal{M}$.
\begin{definition}
	For a metric space ${X}$ with norm $\norm{\cdot}$, a set $\mathcal{M}$ is called an \emph{$\epsilon$-net} for ${X}$ if for every element $x \in {X}$ there exists an element $\tilde{x}\in\mathcal{M}$ such that 
	\be
		\norm{x - \tilde{x}} &\le& \epsilon.
	\ee
\end{definition}
We wish to show that there is a finite $\epsilon$-net for the set of pure states of dimension $d$ under the trace norm.
\begin{lemma}\label{lemma: e-nets}
	For every positive integer $d$ and $0 < \epsilon < 1$, there exists a set of pure state density matrices $\mathcal{M}\equiv \{{\psi_i}\st 1 \le i \le m\}$ such that $m \le \left(\frac{5}{\epsilon}\right)^{2d}$ and for every pure state density matrix $\varphi$, there is some ${\tilde{\varphi}}\in \mathcal{M}$ such that $\trnorm{\varphi - { \tilde\varphi}} < \epsilon$.
\end{lemma}
\begin{proof}
	We give here a version of the proof from $\cite{HLSW04}$. By Lemma \ref{lemma: enet lemma} (p.\pageref{lemma: enet lemma}), to find an $\epsilon$-net of pure density matrices in the trace norm, it suffices to find an $(\epsilon/2)$-net $\mathcal{M'}$ of the set of unit vectors in $\CC^d$, or equivalently $\RR^{2d}$, in the Euclidean norm.
	Let $\mathcal{M'} = \{\ket{\psi_i} \st 1\le i \le m\}$  be a maximal set of unit vectors such that for any $i,j$, $|\ket{\psi_i}-\ket{\psi_j}| \ge \epsilon/2$. Such a set exists by Zorn's Lemma, and is by definition an $(\epsilon/2)$-net for the unit ball in $\RR^{2d}$. We will bound $m$ by a volume argument. Consider the $(\epsilon/4)$-balls around the points of $\mathcal{M}$ as sets in $\RR^{2d}$. The balls are pairwise disjoint, and are all contained in the ball of radius $(1+\epsilon/4)$ centred at the origin. Thus we have 
	\be
		m\frac{\pi^d(\epsilon/4)^{2d}}{\Gamma(d+1)} &\le & \frac{\pi^d(1+\epsilon/4)^{2d}}{\Gamma(d+1)} \\
		\Rightarrow m &\le&  \left(\frac{4+\epsilon}{\epsilon}\right)^{2d}\\
		&\le& \left(\frac{5}{\epsilon}\right)^{2d}.
	\ee
	\end{proof}	

\subsection{Proof of  Theorem \ref{thm: infnorm} }
\begin{proof}
To show the result, we bound the probability that for a random sequence of unitaries, some state is not adequately randomized. That is, for $\{U_i\}$, $i=1, ..., m$, a sequence of $d$ dimensional unitary operator-valued random variables selected according to the Haar measure, we bound
\bes
	\prob{\sup_\rho \infnorm{\frac{1}{m}\sum_{i = 1}^m U_i\rho U^\dagger_i - \frac{\identity}{d}} \ge \frac{\epsilon}{d}}.\label{eqn: tobound}
\ees
By Definition \ref{def: infnorm} (p.\pageref{def: infnorm}), equation (\ref{eqn: tobound}) can be rewritten as follows:
\begin{equation}
	\prob{\sup_{{\sigma},{\varphi}} \left|\frac{1}{m}\sum_{i = 1}^m \tr\left(U_i{\sigma} U^\dagger_i {\varphi} \right) - \frac{1}{d}\right| \ge \frac{\epsilon}{d}}\label{eqn: alt tobound}
\end{equation}
where ${\sigma}$ and ${\varphi}$ can be taken to be pure by the convexity of $|\cdot|$. Now, let $\mathcal{M}$ be an $(\epsilon/2d)$-net of pure states in the trace norm, and for a pure state ${\tau}$, let ${\tilde{\tau}}$ denote an element of $\mathcal{M}$ such that $\trnorm{{\tau} - {\tilde{\tau}}} \le \epsilon/2d$. By Lemma \ref{lemma: e-nets}, we can construct $\mathcal{M}$ such that $|\mathcal{M}|\le \left(\frac{10d}{\epsilon}\right)^{2d}$. Now using the triangle inequality we write 
\bes
		\lefteqn{\prob{\sup_{{\sigma},{\varphi}} \left|\frac{1}{m}\sum_{j = 1}^m \tr\left(U_i{\sigma} U^\dagger_j{\varphi} \right) - \frac{1}{d}\right| \ge \frac{\epsilon}{d}}}\nonumber \\
		&\le& {\rm Pr}\left[\sup_{{\sigma}} \sup_{{\varphi}} \sum_{j=1}^m \frac{1}{m} \left|\tr(U_j\sigma U^\dagger_j \varphi) - \tr(U_j\tilde\sigma U^\dagger_j \tilde\varphi)\right|\right. +\nonumber\\
		&&\qquad\left.\sum_{j=1}^m \frac{1}{m} \left|\tr(U_j\tilde\sigma U^\dagger_j \tilde\varphi)- \frac{1}{d}\right| \ge \frac{\epsilon}{d} \right]\label{eqn: triangle ineq}.
\ees
Applying the triangle inequality to the first term, we have
\be
	\lefteqn{ \left|\tr(U_j\sigma U^\dagger_j \varphi) - \tr(U_j\tilde\sigma U^\dagger_j \tilde\varphi)\right|}\\
	&\le&\left| \tr\left(U_j \sigma U^\dagger_j \varphi\right) - \tr\left(U_j \sigma U^\dagger_j \tilde\varphi \right)\right|
+ \left| \tr\left(U_j \sigma  U^\dagger_j\tilde\varphi\right) - \tr\left(U_j \tilde\sigma U^\dagger_j \tilde\varphi\right)\right|\\
&=& \left|\tr\left(  \left(U_j\sigma U^\dagger_j\right)\left( \varphi -  \tilde\varphi\right)\right)\right| +
\left|\tr\left(  \left(U_j^\dagger\tilde\varphi U_j\right)\left( \sigma -  \tilde\sigma\right)\right)\right| 
\ee
which, as the first term of each of the products in the trace are simply rank 1 projectors is bounded by
\be
\infnorm{\sigma - \tilde\sigma} +
\infnorm{\varphi - \tilde\varphi}.
\ee
As $\sigma - \tilde\sigma$ is either $0$ or rank $2$, and $\mathcal{M}$ is an $(\epsilon/2d)$-net we have
\be
\infnorm{\sigma - \tilde\sigma} +
\infnorm{\varphi - \tilde\varphi} 
&\le&
\frac{1}{2}\trnorm{\sigma - \tilde\sigma} +
\frac{1}{2}\trnorm{\varphi - \tilde\varphi} \\
&\le& \frac{\epsilon}{2d}.
\ee
So we can bound (\ref{eqn: triangle ineq}) by
\bes
\lefteqn{\prob{\sup_{{\sigma},{\varphi}} \left|\frac{1}{m}\sum_{j = 1}^m \tr\left(U_i{\sigma} U^\dagger_j{\varphi} \right) - \frac{1}{d}\right| \ge \frac{\epsilon}{d}}} \nonumber\\&\le&
\prob{\sup_{{\sigma},{\varphi}}\sum_{j=1}^m \frac{1}{m} \left|\tr(U_j\tilde\sigma U^\dagger_j \tilde\varphi)- \frac{1}{d}\right| \ge \frac{\epsilon}{2d}}\label{eq: bound}.
\ees
Now we can replace the the optimization over pure states in (\ref{eq: bound}) by the  maximum over $\mathcal{M}$, and apply a union bound to obtain
\be
	\lefteqn{\prob{\max_{\ket{\tilde\sigma}, \ket{\tilde\varphi} \in \mathcal{M}}  
	\left| \frac{1}{m}\sum_{i=1}^m \tr(U_i \tilde\sigma U^\dagger_i \tilde\varphi) - \frac{1}{d} \right| \ge \frac{\epsilon}{2d}}} \\
	&\le& |\mathcal{M}|^2 \max_{\ket{\tilde\sigma}, \ket{\tilde\varphi} \in \mathcal{M}}   \prob{\left| \frac{1}{m}\sum_{i=1}^m \tr(U_i \tilde\sigma U^\dagger_i \tilde\varphi) - \frac{1}{d} \right| \ge \frac{\epsilon}{2d}}\\
	&\le& \left(\frac{10d}{\epsilon}\right)^{4d} \exp\left(-\frac{Cm\epsilon^2}{4} \right).
\ee
If this probability is bounded away from 1, it means that there is a non-zero probability that every state is randomized by a random selection of $m$ unitaries, and thus there exists a randomizing map using at most $m$ operators. This last is bounded away from $1$ for 
\be
	m &>& \frac{16d}{C\epsilon^2}\ln\left(\frac{10d}{\epsilon}\right)
\ee
and for $\epsilon \ge \frac{10}{d}$ this gives 
\be
	m &>& \frac{134d\ln d }{\epsilon^2}.
\ee
For smaller $\epsilon$, we have $m \ge d^3$, which is worse than the upper bound of $d^2$ achieved by using Pauli (or Heisenberg-Weyl) operators.
\end{proof}

\section{Trace Norm}

Inspired by the previous theorem, we were able to prove a more relevant result (in the sense that the trace norm is generally a more useful measure than the infinity norm), Theorem~\ref{thm: trnorm} (p.\pageref{thm: trnorm}), in the trace norm. 

Our proof is quite similar in structure to that used by Hayden \etal, but differs substantially in the details. We use McDiarmid's inequality \cite{McD89} rather than Cram{\`e}r's Theorem in the large deviation bound, and can use a coarser net (i.e., fewer states) due to working in the trace norm. Before proving Theorem \ref{thm: trnorm}, we introduce McDiarmid's inequality.

\begin{proposition}[McDiarmid's Inequality~\cite{McD89}]\label{thm: mcdiarmid}
Let~$X_1,X_2,\ldots,X_n$ be~$n$ independent random variables,
with~$X_k$ taking values in a set~$A_k$ for each~$k$. Suppose that the
measurable function~$f : \prod_{i = 1}^{n} A_i \rightarrow \reals$ satisfies
\be
\size{f(x) - f(x')} & \le & c_k
\ee
whenever the vectors~$x$ and~$x'$ differ only in the~$k$-th
coordinate. Let~$Y = f(X_1,X_2,\ldots,X_n)$ be the corresponding random
variable. Then for any~$t \ge 0$,
\be
\Pr\left[ Y - \expct(Y) \ge t \right] 
    & \le &  \exp\left(- \frac{2 t^2}{\sum_{i=1}^n c_k^2} \right).
\ee
\end{proposition}

\begin{proof}[Proof of Theorem \ref{thm: trnorm}.]
Once again, we consider a sequence of $m$ unitary operators $\{U_i\}$ chosen independently and uniformly according to the Haar measure on $\UU(d)$ and the corresponding randomizing map
\be
\EE(\rho) &=& \frac{1}{m} \sum_{i = 1}^m U_i\rho U_i^\dagger.
\ee


Fix a pure state $\varphi \in \LL(d)$. Our first step is to bound the expected distance of $\EE(\varphi)$ from the totally mixed state $\identity/d$. Define the random variable $Y_\varphi$ as
\be
	Y_\varphi &=& \trnorm{\EE(\varphi) - \frac{\identity}{d}}.
\ee
By Lemma \ref{lemma: norm relations} (p.\pageref{lemma: norm relations}) we have 
\bes\label{eqn-y}
	Y_\varphi^2 &\le& d \fnorm{\EE(\varphi)-\frac{\identity}{d}}^2
\ees
Then by definition of the $2$-norm (Definition \ref{def: 2-norm}, p.\pageref{def: 2-norm}), 
\bes
\lefteqn{ \fnorm{ \EE(\varphi) - \frac{\identity}{d} }^2 } \nonumber  \\
    & =  & \tr \left( \EE(\varphi) - \frac{\identity}{d} \right)^2 \nonumber \\
    & =  & \tr\,  \EE(\varphi)^2 - \frac{2}{d} \tr \, \EE(\varphi) 
           + \tr\, \frac{\identity}{d^2} \nonumber \\
\label{eqn-f}
    & =  &  \tr\,  \EE(\varphi)^2 - \frac{1}{d}.
\ees
Expanding~$\EE(\varphi)$, we get
\bes
 \tr  \EE(\varphi)^2 
  & = &  \frac{1}{m^2} \sum_i \tr \left( U_i \varphi U_i^\adjoint \right)^2
        + \frac{1}{m^2} \sum_{i \not= j} \tr\left( U_i \varphi U_i^\adjoint 
          U_j \varphi U_j^\adjoint \right) \nonumber \\
\label{eqn-r}
  & = & \frac{1}{m} + \frac{1}{m^2} \sum_{i \not= j} 
          \tr\left( \varphi U_i^\adjoint U_j \varphi U_j^\adjoint U_i \right).
\ees
Here, we have used the fact that~$\tr(\sigma^2) = 1$ for any pure
state density matrix~$\sigma$.

Recall that the unitaries are chosen randomly according to the Haar
measure.
Taking expectation over the random
choice of unitaries, we get
\bes
\lefteqn{ \expct_{\set{U_i}} [ \tr \,  \EE(\varphi)^2 ] }  \label{eqn:expect}\\
  & \le & \frac{1}{m} +  \frac{1}{m^2} \sum_{i \not= j} \expct_{\set{U_i}} 
          \tr\left( \varphi U_i^\adjoint U_j \varphi U_j^\adjoint U_i \right)
          \nonumber \\ 
\label{eqn-er}
  & \le & \frac{1}{m} +  \expct_U
          \tr\left( \varphi U \varphi U^\adjoint \right),
\ees
where the unitary~$U$ has the same distribution as the~$U_i$.  For all
pairs~$i\not=j$, the product~$U_i^\adjoint U_j$ has the same
distribution, that of~$U$, since the underlying sample space~$\UU(d)$
is a group under matrix multiplication.

Note that for a pure
state~$\varphi = \proj{\varphi}$,
\begin{align}
\expct_U  \trace\left( \varphi U \varphi U^\adjoint \right) 
  & =~~ \expct_U \size{ \bra{\varphi}U\ket{\varphi} }^2 &\notag \\
  & =~~ \expct_\psi \size{ \braket{\varphi}{\psi} }^2, &
  \textrm{where } \ket{\psi} \textrm{ is uniformly distributed}\\
\label{eqn-expct}
  & =~~ \frac{1}{d}. & \textrm{By symmetry}
\end{align}

Stringing equations~(\ref{eqn-y}, \ref{eqn-f}, \ref{eqn-r},
\ref{eqn-er}, \ref{eqn-expct}) together, we get
\bes
E\, Y_\varphi
  & \le &  \sqrt{d} \; \expct \fnorm{ \EE(\varphi) - \frac{\identity}{d} }
           \nonumber \\
  & =   & \sqrt{d} \;  \expct \left( \trace\,  \EE(\varphi)^2 - \frac{1}{d}
          \right)^{1/2}
           \nonumber \\
  & \le & \sqrt{d} \; \left( \expct[ \trace\,  \EE(\varphi)^2] - \frac{1}{d}
          \right)^{1/2}, \textrm{~~By Jensen's inequality}
           \nonumber \\
  &  \le & \sqrt{\frac{d}{m}}. \label{eqn-ey}
\ees

We now note that the function~$f_\varphi(U_1,U_2,\ldots,U_m)$ defining the
random variable~$Y_\varphi$ has bounded difference. In
other words, if we replace any one of the unitaries~$U_i$ by another
unitary~$\tilde{U}_i$, the function value changes by a small amount.
Denote the randomizing map given by the modified
sequence
$$(U_1,U_2,\ldots,U_{i-1},\tilde{U}_i,U_{i+1}, \ldots, U_m)$$
by~$\tilde{\EE}$. Then, we have
\begin{align}
\lefteqn{\size{f_\varphi(U_1,U_2,\ldots,\tilde{U}_i,\ldots,U_m)
   - f_\varphi(U_1,U_2,\ldots,U_m)} } \\
  & =~~ \size{ \trnorm{\EE(\varphi) - \frac{\identity}{d} }
        - \trnorm{\tilde{\EE}(\varphi) - \frac{\identity}{d} } } 
        & \notag \\
  & \le~~ \trnorm{ \EE(\varphi) - \tilde{\EE}(\varphi) }, & \textrm{By the
        triangle inequality} \notag \\
  & =~~ \frac{1}{m} \trnorm{ U_i \varphi U_i^\adjoint - \tilde{U}_i \varphi
        \tilde{U}_i^\adjoint } & \notag \\
  & \le~~ \frac{2}{m}. \label{eqn-ck} & 
\end{align}

The McDiarmid bound (Proposition~\ref{thm: mcdiarmid}, along with
equation~(\ref{eqn-ck})) immediately implies that
for any fixed pure state~$\varphi \in \complex^d$,
\be
\Pr[Y_\varphi - \expct\, Y_\varphi \ge \delta] 
& \le & \exp\left( -\frac{\delta^2 m}{2} \right).
\ee
This implies, using our bound from equation~(\ref{eqn-ey}) on the
expected value of~$Y_\varphi$,
\bes
\label{eqn-tailbound}
\Pr[Y_\varphi \ge \delta + \sqrt{d/m}] 
& \le & \exp\left( -\frac{\delta^2 m}{2} \right).
\ees

From Lemma~\ref{lemma: e-nets}, we know that every pure state~$\varphi
\in \complex^d$ is~$\eta$-close in trace norm to a pure 
state~$\tilde{\varphi}$ from a finite set~$\emm$
of size~$\size{\emm} \le \left(\frac{5}{\eta}\right)^{2d}$. By the
triangle inequality, and the unitary invariance of the trace norm~(Lemma~\ref{lemma: unitarily invariant}, p.\pageref{lemma: unitarily invariant}),
it is straighforward to show that~$\size{Y_\varphi -
Y_{\tilde{\varphi}}} \le \eta$. Therefore, if~$Y_\varphi \ge \epsilon$,
then~$Y_{\tilde{\varphi}} \ge \epsilon - \eta$ for some~$\tilde{\varphi} \in
\emm$.

We can now bound the probability that the map~$R$ fails to randomize
some pure state.
\begin{align*}
\lefteqn{ \Pr\left[ \exists \varphi : Y_\varphi > \epsilon \right] } \\
  & \le~~ \Pr\left[ \exists \tilde{\varphi} \in \emm :
          Y_{\tilde{\varphi}} > \epsilon - \eta \right]
          & \textrm{From the discussion above} \\
  & \le~~ \size{\emm} \cdot \Pr \left[ Y_{\tilde{\varphi}} > \epsilon - \eta \right]
          & \textrm{By the union bound} \\
  & \le~~ \left(\frac{5}{\eta}\right)^{2d} 
          \exp\left( -(\epsilon - \eta - \sqrt{d/m}\,)^2 \frac{m}{2} \right)
          & \textrm{By equation~(\ref{eqn-tailbound})} \\
  & \le~~ e^{-d/2}, & 
\end{align*}
if~$\eta$ is chosen to be at most~$\epsilon/3$, and~$m$ at least
\[
\frac{37d}{\epsilon^2} \ln \left( \frac{15}{\epsilon} \right).
\]
Thus, the overwhelming majority of random sequences of~$m$ unitaries selected according to the Haar measure are randomizing to within~$\epsilon$, with respect to the trace norm.

\end{proof}

\subsection{Using Pauli Operators instead of the Haar Measure}
The only place we used information about the distribution of the operators chosen was in calculating (\ref{eqn:expect}). If instead of taking unitaries distributed according to the Haar measure, we instead select only among Pauli matrices, the same result holds. 

Let $d = 2^n$ for some integer $m$, and consider 
\[
\EE(\varphi) = \frac{1}{m}\sum_{i=1}^m P_i \varphi P_i^\dagger
\]
where the $P_i$ are uniformly distributed Pauli matrices. Observe that the product of two uniformly distributed random Paulis is, up to sign, another uniformly distributed random Pauli. Thus we can say as before:
\[
\expct_{\{P_i\}}[\trace\,\EE(\varphi)^2] \le \frac{1}{m} + \expct_P\left[\trace\left(\varphi P \varphi P^\dagger\right)\right]
\]
and we have to calculate 
\bes
\expct_P |\trace(P \varphi)|^2&=&\frac{1}{d^2}\sum_{i=1}^{d^2}|\trace(P_i \varphi)|^2\notag\\
&=&\frac{1}{d}\sum_{i=1}^{d^2}\left|\frac{\trace(P_i \varphi)}{\sqrt{d}}\right|^2 \notag\\
&=&\frac{1}{d}{\fnorm{\varphi}^2}\label{eqn:orthonormal}\\
&=&\frac{1}{d}\notag
\ees

Where (\ref{eqn:orthonormal}) is because the set of Pauli matrices forms an orthogonal basis for for the space of unitary operators under the inner product
\[
\langle U, V \rangle ={\trace (U^\dagger V )}
\]
(Lemma \ref{lemma: paulis orthogonal basis}, p.\pageref{lemma: paulis orthogonal basis}), and each unitary has norm $\fnorm{U} = \sqrt{d}$.

Thus, when we are working with spaces of appropriate dimension we can choose the unitaries in Theorem \ref{thm: trnorm} uniformly among the Pauli operators. This is important, as the Pauli operators are efficiently describable, and (presumably) relatively easily constructible.


\clearpage
\thispagestyle{empty}
\cleardoublepage

\chapter{Explicit Constructions of Approximate Randomizing Maps}\label{chap: explicit}
\markright{Explicit Constructions of Approximate Randomizing Maps}

\section{Previous Work}
In Chapter \ref{chap: probabilistic}, we showed that there exist $\epsilon$-randomizing maps on $n$ qubits that require at most $n + \bigoh(\log(1/\epsilon))$ bits of key. In this Chapter, we give explicit constructions of maps that meet this bound. 

The first explicit maps were exhibited by Ambainis and Smith \cite{AS04}, and make heavy use of the constructions of $\epsilon$-biased spaces of Alon \etal~\cite{AGHP92} and Alon, Bruck \etal~\cite{ABNNR92}. For an introduction to $\epsilon$-biased spaces, see Appendix \ref{chap: small bias}.  In \cite{AS04}, Ambainis and Smith give three constructions of $\epsilon$-randomizing maps. The first uses $n + 2\log n + 2\log(1/\epsilon) + \bigoh(1)$ bits of key, the second $n + 2\log(1/\epsilon)$ bits of key, plus $2n$ additional classical bits of communication, and the third $n + \min\{2\log n + 2\log(1/\epsilon), \log n + 3 \log(1/\epsilon)\}+\bigoh(1)$ bits of key. The first two schemes use only Pauli operators for the maps, and the third uses Heisenberg-Weyl operators in $d$ dimensions. All are constructible in polynomial time. 

Our contribution is an improvement over the schemes of \cite{AS04}; we provide a scheme using $n + 2\log(1/\epsilon) + \bigoh(1)$ bits of key, but without additional classical communication overhead. Our construction is based on one of the constructions of Alon, Goldreich, H{aa}stad and Peralta \cite{AGHP92} and a theorem of Ambainis and Smith connecting the existence of small bias spaces with $\epsilon$-randomizing maps.

\section{Construction of $\epsilon$-Randomizing Map}

Ambainis~and~Smith~\cite{AS04} prove the following theorem.
\begin{theorem}\label{thm: small bias implies randomizing}
Given a $\delta$-biased space $\ESS\subset\{0,1\}^n \times \{0,1\}^n$, the map
\begin{eqnarray*}
	\EE(\rho) = \frac{1}{|S|}\sum_{(a,b)\in \ESS} X^a Z^b \rho Z^b X^a
\end{eqnarray*}
is $\epsilon$-randomizing on $n$-qubit states for any $\epsilon$ such that 
\begin{eqnarray*}
	\epsilon &\ge& 2^{n/2}\delta
\end{eqnarray*}
\end{theorem}
We include a version of their proof, as it is critical to our construction.
\begin{proof}
	Let $\ESS\subset \{0,1\}^{2n}$ be a $\delta$-biased space. We interpret elements of $\ESS$ as a concatenated pair of $n$-bit strings $(a,b)$, and let
	\be
		\EE(\rho) &=& \sum_{(a,b)\in \ESS}\frac{1}{|\ESS|} X^a Z^b \rho Z^b X^a.
	\ee
Since we have 
\be
	\fnorm{\rho - \frac{\identity}{2^n}}^2 &=& \tr\left(\left(\rho- \frac{\identity}{2^n}\right)\left(\rho- \frac{\identity}{2^n}\right)^\dagger\right)
\ee
and as $\EE(\rho)$ is a density matrix, it is Hermitian and has trace $1$, $\left(\EE(\rho) - \frac{\identity}{2^n}\right)^\dagger = \EE(\rho) - \frac{\identity}{2^n}$, using linearity of trace we obtain
\bes
	\fnorm{\EE(\rho) - \frac{\identity}{2^n}}^2 &=& 
			\tr((\EE(\rho))^2) - 2\frac{\tr(\EE(\rho)\identity)}{2^n} + \frac{\tr(\identity^2)}{2^{2n}}\nn\\
		&=& \tr(\EE(\rho^2)) - \frac{1}{2^n}.\label{eqn: 2norm - id}
\ees
We evaluate $\tr(\EE(\rho^2))$ by expressing it in the Pauli basis.
We can express $\rho$ in the Pauli basis as
\be
	\rho = \frac{1}{2^n} \sum_{u,v\in \{0,1\}^n}\alpha_{u,v} X^uZ^v,
\ee 
with $\alpha_{u,v} = \tr(Z^vX^u\rho)$, and obtain
\bes
	\tr(\EE(\rho)^2) &=& \tr\left(\left(\frac{1}{|\ESS|}\sum_{(a,b)\in \ESS}X^aZ^b\left(\frac{1}{2^n}\sum_{u,v \in \{0,1\}^n} \alpha_{u,v}X^uZ^v\right)Z^bX^a\right)^2\right) \nn\\
	&=& \tr\left(\left( \frac{1}{|\ESS|} \sum_{u,v \in \{0,1\}^n}\frac{1}{2^n}\sum_{(a,b)\in \ESS}  (-1)^{a\cdot v + b \cdot u}\alpha_{u,v}X^uZ^v\right)^2\right) \label{eqn: commutation}\\
	&=& \tr\left(\left(\frac{1}{2^n}\sum_{u,v}\expct_{(a,b)\in \ESS}\left[(-1)^{a\cdot v + b\cdot u}\right] \alpha_{u,v}\right)^2\right)\label{eqn: linearity of expct}\\
	&=& \frac{|\alpha_{0,0}|^2}{2^n}+\frac{1}{2^n}\sum_{(u,v) \neq (0,0) } \left| \expct_{(a,b)\in \ESS}\left[(-1)^{a\cdot v + b\cdot u}\right]\alpha_{u,v}\right|^2\label{eqn: alt def of fnorm}
\ees
where (\ref{eqn: commutation}) comes from the commutation relation between Pauli operators,   (\ref{eqn: linearity of expct}) follows by the linearity of expectation, and Equation (\ref{eqn: alt def of fnorm}) is  from (\ref{eqn: fnorm def}), p.\pageref{eqn: fnorm def}. Then as $\ESS$ is $\delta$-biased, have have
\be
	\tr(\EE(\rho)^2)&\le& \frac{1}{2^n} + \frac{2^n-1}{2^n}\left(\delta^2 \tr(\rho^2)\right)
\ee
Setting $\delta = \frac{\epsilon}{2^{n/2}}$ and using (\ref{eqn: 2norm - id}), as $\fnorm{\rho}\le 1$, we have
\be
	\fnorm{\EE(\rho) - \frac{\identity}{2^n}}^2 &\le& \frac{\epsilon^2}{2^n}
\ee
and by Equation (\ref{eqn: tr<2}) in Lemma \ref{lemma: norm relations} (p.\pageref{lemma: norm relations}),
\be
	\trnorm{\EE(\rho)-\frac{\identity}{2^n}} &\le & \epsilon.
\ee
\end{proof}

We can now use this construction to prove Theorem~\ref{thm: construction} (page \pageref{thm: construction}), i.e., that there exists an $\epsilon$-randomizing map using $n+\log(1/\epsilon^2)$ bits of private key.
\begin{proof}[Proof of Theorem \ref{thm: construction}]
Using  the third $\epsilon$-biased space construction of Alon~\etal~\cite{AGHP92}~(Section \ref{sec: small bias construction}), with bias $\delta = \frac{\epsilon}{2^{n/2}}$ and $rs=2n$, Proposition \ref{prop:  small-bias} gives us
\be
	\frac{\epsilon}{2^{n/2}} &\le& \frac{s-1}{2^r} 
\ee
substituting and taking logs, we obtain
\be
	\log(\epsilon) - \frac{n}{2} &\le& \log\left(\frac{2n}{r}-1\right) - r 
\ee
and by the monotonicity of $\log$,
\bes
	r+\log r &\le& \frac{n}{2} + \log{n} - \log{\epsilon} + 1\label{eqn: key bits needed}. 
\ees
Setting $2r = n + \log\left(\frac{1}{\epsilon^2}\right)$, we have 
\be
	r + \log r &=& \frac{n}{2} + \log\left(\frac{1}{\epsilon}\right) + 
			\log\left( \frac{n}{2} + \log\left(\frac{1}{\epsilon}\right)\right).
\ee
Again, applying monotonicity of log, we have (\ref{eqn: key bits needed}) holds for $\epsilon\ge 2^{-n/2}$, and as we need $2r$ bits to specify a string in $\ESS$, we can use this many bits of key.

For $\epsilon \le 2^{-n/2}$, we can simply use the Quantum One Time Pad (Theorem~\ref{thm:egpqc}, p.\pageref{thm:egpqc}).
\end{proof}

This scheme has several advantages. A circuit for implementing the scheme is also simple, requiring only classical control and one-qubit Pauli gates. No communication outside the $n$-qubit message is necessary. From the length of the received message ($n$ qubits), the recipient can in polynomial time deterministically obtain a primitive polynomial of degree $r$ over $GF(2)$ (using for example the results of Shoup \cite{S89}). As the process is deterministic, both the sender and the receiver will arrive at the same representation of $GF(2^r)$. Given a particular representation, computing the key requires $2r$ multiplications in the field, each taking $O(r^2)$ bit operations, and $2r$ binary inner products, each costing $O(r)$ bit operations. Thus one can compute the key to use in $O(r^3)<O(n^3)$ bit operations. While this is larger than we might like, it is not too unwieldy. As such, this scheme seems to be quite practical and suitable for real use, should the need arise.

As shown by Kerenidis and Nagaj \cite{KN05}, Theorem \ref{thm: small bias implies randomizing} does not hold in the $\infty$-norm. The probabilistic results of Hayden \etal~\cite{HLSW04} remain the best known in the $\infty$-norm.

\clearpage
\thispagestyle{empty}
\cleardoublepage

\chapter{Lower Bounds}\label{chap: lower bounds}
\markright{Lower Bounds}

We have done work towards finding a tight lower bound for the amount of classical entropy needed to $\epsilon$-randomize an arbitrary quantum state, and have shown several results. Prior to this work, the best bound known to the author was $\log d-\bigoh(1)$ bits of randomness required to randomize quantum states of dimension $d$ (we give a proof in Section \ref{sec: naive bound}). This is very unsatisfactory, as we know that for $\epsilon=0$, $2\log d$ bits are necessary, while for $\epsilon > 0$, $\log d + \bigoh(\log(1/\epsilon))$ are known to be sufficient. Thus the independence of the known bound from $\epsilon$ is troubling. 

For the purposes of this chapter, we consider $\epsilon$-randomizing maps  on pure states of $n$ qubits (i.e., $d = 2^n$) with the property 
\be
\EE(\rho) &=& \frac{1}{|\ESS|}\sum_{U \in \ESS}^m U \rho U^\dagger
\ee
where each $U \in \ESS$ is a tensor product of single qubit Pauli operators. We have demonstrated (Sections \ref{sec: weak bound}, \ref{sec: weak alt proof}) a weak lower bound on $|\ESS|$ that is dependent on $\epsilon$ in the trace norm. Our bound is of the form $|\ESS| \ge \Omega\left(\frac{1}{\epsilon^2}\right)$. This is stronger than the na{\"i}ve bound for $\epsilon < 2^{-n/2}$, but is clearly not tight. We have also developed conditions that the set $\ESS$ of operators must satisfy that we hope can be used to prove a stronger lower bound. 

We conjecture that the upper bounds established by our protocol (Theorem \ref{thm: construction}, p.\pageref{thm: construction}) are in fact tight.

\begin{conjecture}
Let $\{U_1, ..., U_m\}$ be a set of unitary operators and $\EE(\rho) = \frac{1}{m}\sum_{j=1}^m U_j \rho U^\dagger_j$. Then if $\EE$ is $\epsilon$-randomizing on $\CC^d$, $m \ge d/\epsilon^c$ for some constant $c$.
\end{conjecture}

\section{Na{\"i}ve Lower Bound}\label{sec: naive bound}
The following bound  on the minimum number of unitary operators needed for an $\epsilon$-randomizing map employs a simple rank argument. It shows that for an $n$-qubit state, at least $n - o(1)$ bits of key are required.

\begin{theorem}\label{thm: naive bound}
Let $\EE: \mathcal{L}(d)\rightarrow\mathcal{L}(d)$ be an $\epsilon$-randomizing map such that there exists a sequence of unitary operators $\{U_j\}_{j=1,..., m}$ with 
$\EE(\rho) = \frac{1}{m}\sum_{j=1}^m U_j \rho U^\dagger_j$,
then $m \ge d(1-\epsilon/2)$.
\end{theorem}
\begin{proof}
Note that $\EE(\rho)$ is a normal operator and thus there exists some unitary $V_\rho$ such that $V_\rho\EE(\rho)V^\dagger_\rho$ is diagonal, and as the trace norm is unitarily invariant (Lemma \ref{lemma: unitarily invariant}, p.\pageref{lemma: unitarily invariant})
\be
	\trnorm{V_\rho \EE(\rho) V^\dagger_\rho - \frac{\identity}{d}} 
	&=& \trnorm{\EE(\rho) - \frac{\identity}{d}}.
\ee
Then, as $\EE(\rho)$ has rank at most $m$, and as $\EE$ is $\epsilon$-randomizing
\be
	2\left(1 - \frac{m}{d}\right) 
	\le& \trnorm{V_\rho \EE(\rho) V^\dagger_\rho - \frac{\identity}{d}} &\le \epsilon\\
	 \Leftrightarrow&m\ge d\left(1 - \frac{\epsilon}{2}\right).
\ee
\end{proof}

\section{Weak Lower Bound in Trace Norm}\label{sec: weak bound}
We wish to demonstrate a lower bound on the number of operators needed to $\epsilon$-randomize an arbitrary quantum state. To begin with, we consider a restricted class of randomizing operators: those formed by taking only Pauli operators on $n$ qubits. In this section, we prove  Theorem \ref{thm: weak bound}, p.\pageref{thm: weak bound}, connecting the existence of approximately randomizing sets with the existence of small-bias spaces.

Note first that in order to randomize all states, $\EE$ must clearly randomize the Pauli eigenvectors. We will develop our bound by considering the results of $\EE$ on these states. 

For an $n$-bit string $c$, define $\proj{c} = \bigotimes_{j = 1}^n \proj{c_j}$. Consider the action of $\EE$ on the state $\ketbra{0}{0}^{\otimes n}$. We have
\be
	\EE(\proj{0}^{\otimes n}) &=& \frac{1}{|\ESS|}\sum_{(a,b)\in\ESS} X^aZ^b \proj{0}^{\otimes n}Z^bX^a \\
	&=& \frac{1}{|\ESS|}\sum_{(a,b)\in\ESS}  \bigotimes_{j=1}^n X^{a_j}Z^{b_j}  \ketbra{0}{0} Z^{b_j}X^{a_j} \\
	&=& \frac{1}{|\ESS|}\sum_{(a,b)\in\ESS}  \bigotimes_{j=1}^n \ketbra{a_j}{a_j} \\
	&=& \frac{1}{|\ESS|}\sum_{(a,b)\in\ESS} \proj{a}
\ee
By replacing the initial state $\proj{0}$ with $\proj{x}$ for some $n$-bit string $x$, the final equation becomes 
\bes
\EE(\proj{k})&=& \frac{1}{|\ESS|}\sum_{(a,b)\in\ESS} \proj{a\oplus x}\label{eqn: randomized Z-evec}
\ees
and as (\ref{eqn: randomized Z-evec}) is diagonal, with $x$ corresponding to a re-ordering of the rows, we have 
\be
	\trnorm{\EE(\proj{x}) - \frac{\identity}{2^n}} &=&\trnorm{\EE(\proj{0}) - \frac{\identity}{2^n}}.
\ee
This demonstrates that it is not important which particular string we start with, only that it is in the same basis.
If instead of considering the $Z$ eigenvectors, we consider $X$ and $Y$, we obtain very similar results.
\be
	\EE(\proj{+}^{\otimes n})
		&=& \frac{1}{|\ESS|}\sum_{(a,b)\in\ESS} X^a Z^b \proj{+}^{\otimes n} Z^b X^a \\
		&=& \frac{1}{|\ESS|}\sum_{(a,b)\in\ESS} Z^b \proj{+}^{\otimes n}  Z^b\\
		&=& \frac{1}{|\ESS|}\sum_{(a,b)\in\ESS} \bigotimes_{j=1}^n \proj{(-1)^{b_j}} \\
		&=&  \frac{1}{|\ESS|}\sum_{(a,b)\in\ESS} H^{\otimes n} \proj{b} H^{\otimes n} 
\ee
and 
\be
	\EE(\proj{+i}^{\otimes n})
		&=& \frac{1}{|\ESS|}\sum_{(a,b)\in\ESS} X^a Z^b \proj{+i}^{\otimes n} Z^b X^a \\
		&=& \frac{1}{|\ESS|}\sum_{(a,b)\in\ESS} \bigotimes_{j=1}^n \proj{(-1)^{a_j \oplus b_j}i} \\
		&=&  \frac{1}{|\ESS|}\sum_{(a,b)\in\ESS} (PH)^{\otimes n} \proj{a\oplus b} (HP)^{\otimes n} 
\ee
where $P = \left(\begin{array}{cc}1&0\\0&i\end{array}\right)$ and the same remark we made with respect to changing $0$ to $1$ holds {\it mutatis mutandis} for $\{+,-\}$ and $\{+i, -i\}$. As we have seen above, the ``sign" on each qubit is immaterial and we need only concern ourselves with the basis ($\{0,1\} \equiv Z$, $\{+,-\} \equiv X$ or $\{+i, -i\} \equiv Y$) that each qubit is drawn from. We formalize this notion here.

A single Pauli eigenvector on $n$ qubits defines an orthonormal basis for the space of states on $n$ qubits as follows. 
\begin{lemma} \label{lemma: eigenvector defines basis}
Given a Pauli eigenvector of dimension $2^n$, there is an orthogonal basis of Pauli eigenvectors for $\CC^d$ including that eigenvector.
\end{lemma}
\begin{proof}
Let $w = w_1, w_2, ..., w_n$, and $V = V_1\otimes...\otimes V_n$ where each $V_i\in\{X,Y,Z\}$ and $w_i \in \{0,1\}$. Define $\proj{w,V} = \bigotimes_{j=1}^n \frac{\identity + (-1)^{w_n} V_n}{2}$ (a tensor product of Pauli eigenvectors). For each $c \neq c' \in \{0,1\}^n$, observe that $\proj{c, V}$ and $\proj{c',V}$ are orthogonal (as the inner product commutes with the tensor product), and that as there are $2^n$ such states, they form a basis for $\CC^{2^n}$. Thus the string $V$ can be taken to define a basis in this manner.
\end{proof}
These bases form equivalence classes of Pauli eigenvectors, $\ket{w,V} \cong \ket{w', V}$ for every pair of strings $b$ and $b'$ and as we have seen above 
\bes
	\trnorm{\EE(\proj{w,V}) - \frac{\identity}{2^n}} &=& \trnorm{\EE(\proj{w',V}) - \frac{\identity}{2^n}}.
	\label{eqn: eqclass}
\ees

As these operators are all diagonal, we can make the following simplification. For any $V \in \{X,Y,Z\}^{\otimes n}$ define $\sigma_V : \ESS \rightarrow \{0,1\}^n$ as follows:
\be
	\sigma_V((a,b))_j &=& 
		\begin{cases}
			a_j & \text{if } V_j = Z \\
			b_j & \text{if } V_j = X \\
			a_j\oplus b_j & \text{if } V_j = Y
		\end{cases}
\ee
Now for any Pauli eigenstate $\ket{w,V}$, we can write $X^aZ^b\ket{w,V} = \ket{\sigma_V((a,b))\oplus w, V}$, and using (\ref{eqn: eqclass}), we have
\be
	\trnorm{\EE(\proj{w,V}) - \frac{\identity}{2^n}} &=& \trnorm{ \frac{1}{|\ESS|}\sum_{(a,b)\in\ESS} \proj{\sigma_V((a,b))\oplus w,V} - \frac{\identity}{2^n}}\\
	 &=& \trnorm{ \frac{1}{|\ESS|}\sum_{(a,b)\in\ESS} \proj{\sigma_V((a,b)), V} - \frac{\identity}{2^n}}
\ee
as each of these is diagonal in the $V$ basis, this is the same as taking the variation distance from the uniform distribution:
\be
		&=& \sum_{k \in \{0,1\}^n} \left| \frac{\left|\sigma^{-1}_V(k)\right|}{|\ESS|} -  \frac{1}{2^n}\right|
\ee
where $\sigma_V^{-1}(k)$ is the preimage of $k\in\{0,1\}^n$. Then the $\epsilon$-randomizing condition is a requirement that the distribution $\sigma_V(\ESS)$ is ``nearly uniform" for every basis $V$. More precisely, define the distribution $D_V$ on $\{0,1\}^n$ as $D_V(x) = \frac{1}{|\ESS|} \left|\sigma_V^{-1}(x)\right|$. For $\EE$ to be $\epsilon$-randomizing, it is necessary that
\be
	\max_V\sum_{x \in \{0,1\}^n} \left| D_V(x) - \frac{1}{2^n} \right | \le \epsilon.
\ee

In fact, the following lemma shows that we can take the maximum over all strings, rather than just valid bases. This requires that we generalize the definition $D_V$.

\begin{lemma}
Let $\ESS$ be a set of strings such that $\EE$ is $\epsilon$-randomizing on all $n$-qubit Pauli eigenstates. For every string $W \in \{00,01,10,11\}^n$, let $k_W$ be the number of times that $00$ occurs in $W$. Then
\be
	\max_W\sum_{x \in \{0,1\}^{n-k_W}} \left| D_W(x) - \frac{1}{2^{n-k_W}} \right | \le \epsilon.
\ee
\end{lemma}

\begin{proof}
This follows as $D_W$ is simply a marginal distribution of some $D_V$, where $V$ is as above.


\end{proof}

Each $W$ can be taken as a characteristic vector selecting a subset of bit positions from $1,..., 2n$ in $\ESS$, where the $j$th bit position is selected by $W$ if $W_j=1$. Each $D_V$ is $\epsilon$-away from independence, thus 
as discussed in section \ref{sec: smallbias defns}, for 
any subset  of the random variables consisting of the bit positions of $\ESS$, the $XOR$ of those bit positions has bias at most $\epsilon$, and thus $\ESS$ is $\epsilon$-biased. Then by Theorem \ref{thm: e-bias lower bound} (p.\pageref{thm: e-bias lower bound}), $|\ESS|\ge \Omega\left(\frac{n}{\epsilon^2\log(1/\epsilon)}\right)$, which establishes Theorem \ref{thm: weak bound}.

While this result is not tight, and in fact for $\epsilon \ge 2^{-n/2}$ is weaker than the naive bound, it is interesting in that it does depend on $\epsilon$, and it is the best known lower bound for $2^{-n}\le\epsilon\le2^{-n/2}$. 

\section{An Alternative Proof}\label{sec: weak alt proof}
Subsequent to finding the previous result, we examined the effect of applying $\EE$ to other types of states in the hopes of finding more conditions on $\ESS$. By  considering so called ``cat" states rather than tensors of Pauli eigenvectors, we developed the following alternative proof of Theorem \ref{thm: weak bound}.

In general, a cat state is a pure quantum state $\ket{\phi}$ such that if any subsystem is traced out one is left with a random choice of two orthogonal states. For our purposes, we will consider only those cat states of the form $\frac{1}{\sqrt{2}}\ket{\varphi}+\ket{\tilde\varphi}$ where $\ket{\varphi}$ is a Pauli eigenvector $\ket{w,V}$, and $\ket{\tilde{\varphi}}$ is $\ket{\overline{w}, V}$, where $\overline{w}$ is the complement of $w$.

Similarly to the previous proof, we can consider the result of an $\epsilon$-randomizing map on a cat state. First, consider $\ket{\phi} = \frac{1}{\sqrt{2}}(\ket{0^n} + \ket{1^n})$. Applying $X^a Z^b$ to $\ket{\phi}$ yields the state 
\begin{eqnarray}
	\frac{1}{\sqrt{2}}(\ket{a} + (-1)^{|b|} \ket{\bar{a}}).\label{eqn: X^aZ^b phi}
\end{eqnarray}
where $|b|$ is the Hamming weight, or number of $1$s in $b$.

 If a set of strings $\ESS$ is to have the $\epsilon$-randomizing property, then we must have
\be
	\trnorm{\frac{1}{|\ESS|}\sum_{(a,b)\in \ESS} X^aZ^b \ketbra{\phi}{\phi} Z^bX^a - \frac{\identity}{2^n}}
	&\le& \epsilon
\ee
As $\{\frac{1}{\sqrt{2}}\left(\ket{a} \pm \ket{\overline{a}}\right)\}$ forms a basis for state space, this means that we must have that $\bias(|b|) \le \epsilon$. To justify this claim, observe that if $\bias{|b|} \ge \epsilon$, with say, $\prob{|b| = 0 \mod 2}  = p > \frac{1+\epsilon}{2}$, then the projection onto the subspace spanned by the ``plus" states will have probability $p$, and onto the minus states $1-p$, and the bias of this variable will be at least $\epsilon$.

In fact, we get the same result if instead of using $0^n$ and $1^n$, we use an arbitrary $\{0,1\}$ string $w$ and its complement $\bar{w}$, the result is  $\frac{1}{\sqrt{2}}(\ket{a\oplus w} + (-1)^{|b|} \ket{a \oplus \bar{w}})$. So, as previously, we ignore the exact state, and consider only the basis it is chosen from. That is,  instead of choosing $w\in\{0,1\}^n$, we choose it in $\{X,Y,Z\}^n$. Thus our example state, and the restriction on $\bias(|b|)$, was the result of choosing $w = Z^n$. If we choose the string $w = X^n$, we similarly observe that $\bias(a) \le \epsilon$ by conjugating with Hadamards, and invoking the unitary invariance of the trace norm (Lemma \ref{lemma: unitarily invariant}, p.\pageref{lemma: unitarily invariant}) and similarly $\bias(|a\oplus b|)$ is bounded by considering $w = Y^n$. As before, we can extend $w$ to include $\identity$. Having an $\identity$ in the $j$th position of $w$ means that we do not test the value of either $a_j$ or $b_j$. It corresponds to randomizing a state that has the $j$th qubit in the totally mixed state. 

To prove the result, we must show that for any subset of the bits of $(a,b)$, ie, for any $W\subset[2n]$, the bias of $W$ in $\ESS$ is at most $\epsilon$. We identify $X = 10$, $Y= 11$, $Z = 01$, and $\identity = 00$, and then for each possible $W$, we can construct an appropriate state $w$, and claim that if $\EE$ is $\epsilon$-randomizing, the subset of bit positions corresponding to $W$ must be unbiased, and thus the result follows.

This proof demonstrates an interesting point, while for example in considering the state (\ref{eqn: X^aZ^b phi}) we only focus on the bias the bits of $b$, we completely ignore the stronger condition that the strings $a$ must be nearly independent. In developing stronger bounds, this may be of interest.
%

	
\section{Conditions For $\epsilon$-Randomizing Maps}	

By examining the action of $\EE$ on certain classes of states, we can discover conditions on the maps in $\ESS$ that may eventually lead to a tigher lower bound.

\subsection{Subspace States}
We consider the results of applying Pauli maps to uniform superpositions over linear subspaces of $\ZZ_2^n$. These spaces correspond to codewords for linear codes and this proof requires some familiarity with the same. For an introduction to the subject covering all the required material, see $\cite{NC00}$. 

Let $\{w_1, ..., w_k\}$ be linearly independent vectors in $\ZZ_2^n$. Let $W = \text{span}\{w_1,..., w_k\}$ over $\ZZ_2$. Then a uniform superposition over $W$, which we represent $\ket{W}$ is
\be
	\ket{W} &=& \frac{1}{\sqrt{|W|}} \sum_{w \in W} \ket{w}.
\ee
Also define $W^\perp$ to be the dual space of $W$, that is, 
\be
	W^\perp &=& \{v \st w\cdot v = 0 \quad \forall w \in W\}.
\ee
If $\tilde{W}$ is the linear operator with rows $w_1, ..., w_k$, then we have $W^\perp$ is the kernel of the linear map $\tilde{W}$, and hence $\dim(W) + \dim(W^\perp) = n$, and thus we can find linearly independent $w_{k+1}, ..., w_{n}$ such that $W^\perp = \mathrm{span}\{w_{k+1}, ..., w_{n}\}$.

\begin{lemma}\label{lemma: linear space states}
	Let $X^aZ^b$ be a Pauli operator. Then 
	\be
		\bra{W}X^aZ^b\ket{W} &=& \begin{cases} 
			1 & \text{if $a\in W$ and $b\in W^\perp$} \\
			0 & \text{otherwise}
			\end{cases}.
	\ee
\end{lemma}
\begin{proof}
	If $a\in W$, then applying $X^a$ to $\ket{W}$ will simply permute the elements of $W$. If $a\notin W$, then for each $w\in W$, $a\oplus w \notin W$, and hence the inner product will have disjoint elements on the left and right, and thus every pair will annihilate. For $Z^b$, we have 
	\be
		Z^b\ket{W} &=&  \frac{1}{\sqrt{|W|}} \sum_{w \in W} (-1)^{b\cdot w}\ket{w}.
	\ee
Thus, if $b \in W^\perp$, $Z^b\ket{W} = \ket{W}$. However, if $b\notin W^\perp$, then there is some basis of $W$ $\{v_1, ..., v_k\}$ such that $b\cdot v_1 = 1$ and $b \cdot v_j = 0$ for each $j = 2, ..., k$. Thus by linearity, $b\cdot w = 1$ for exactly half of the strings in $W$, and hence $\bra{W}Z^b\ket{W} = 0$. Finally, we observe that after applying $Z^b$, if $a\notin W$, the signs on each element of the superposition are irrelevant, and that if $a \in W$, we simply change which elements have $-1$ factors, and not the sum. Thus $\bra{W}X^aZ^b\ket{W} = 1$ if $a\in W$ and $b \in W^\perp$, and $0$ otherwise.
\end{proof}
This immediately yields the following corollary.
\begin{corollary}
Given strings $a, a', b, b'$ and $\ket{W}$ as above, 
\be
	\left|\left(\bra{W}Z^{b'}X^{a'}\right)\left(X^aZ^b\ket{W}\right)\right| &=&
	\begin{cases}
		1 & \text{if $a+a' \in W$ and $b+b' \in W^\perp$} \\
		0 & \text{otherwise}.
	\end{cases}
\ee 
\end{corollary}
As $W$ and $W^\perp$ are normal subgroups of $\ZZ_2^n$, we can define the coset map 
\be
(a,b) \mapsto (a+W, b+W^\perp),
\ee
and observe that $X^aZ^b\ket{W}$ is constant on pairs of cosets, and orthogonal for different pairs of cosets. Further, as $|\ZZ_2^n/W| \times |\ZZ_2^n/W^\perp| = 2^{n-k} 2^{k}$, we have that these coset states form an orthogonal basis for the state space on $n$-qubits.

We can use the fact that $W$ is a linear subspace to find explicit coset representatives by exploiting the parity check matrix for $W$. Let $H$ be the matrix with rows $w_{k+1}, ..., w_{n}$, that is, the basis vectors of $W^\perp$. Clearly $Hw = 0$ for every $w\in W$. $H$ is called the parity check matrix for the code $W$.  Furthermore, $H^T$ is the generator matrix for the dual subspace, $W^\perp$. Similarly, $G$, the matrix with columns $w_1, ... w_k$ is the generator matrix for $W$, and $G^T$ the parity check matrix for $W^\perp$. Thus the map 
\be
	(a,b) &\mapsto& (Ha, G^T b)
\ee
gives us an explicit representative for the basis element given by $X^aZ^b\ket{W}$. We will write $\tilde{a}$ and $\tilde{b}$ for the cosets of $a$ and $b$ respectively, and denote basis elements by 
\be
\ket{\tilde{a},\tilde{b}} &=& X^aZ^b\ket{W}.
\ee
Now, we can consider the result of applying $\EE$ to $\ket{W}$.
\be
	\EE(\proj{W}) &=& \frac{1}{|\ESS|}\sum_{(a,b)\in\ESS} X^aZ^b\proj{W}Z^bX^a \\
	&=&  \frac{1}{|\ESS|}\sum_{(a,b)\in\ESS} \proj{\tilde{a},\tilde{b}}
\ee
This now tells us that in order for $\EE$ to be $\epsilon$-randomizing, we must have $\ket{\tilde{a},\tilde{b}}$ at most $\epsilon$-away from independent for every linear subspace $W$. This is a different condition from those established earlier, as it tells us about the action of $\EE$ on a different class of states (this class includes CSS code states).

\subsection{Stabilizer States}

We can generalize the results of the previous section from subspace states to general stabilizer states. Stabilizer states are a large and important class of quantum states in quantum computation introduced by Gottesman \cite{G97}. For an introduction, see either Daniel Gottesman's PhD thesis \cite{G97} or Chapter 10.5 of \cite{NC00}. We include all the relevant results and definitions in Appendix \ref{chap: stabilizers}.
%

Let $\ket{\psi}$ be a stabilizer state, with $\stab(\ket{\psi}) = G = \langle g_1, ..., g_n \rangle$. Then $G < \PP_n$ (here, $\PP_n$ is the set of $n$-fold tensors of Pauli operators, see Section \ref{sec: pauli group}, p.\pageref{sec: pauli group}), and every element $P\in\PP_n$ either commutes with every element of $G$, or there is some set of generators $g_1,...,g_j$  such that only $g_1$  doesn't commute with $P$ (Lemma \ref{lemma: one generator}, p.\pageref{lemma: one generator}).

Consider a stabilizer state $\ket{\psi}$ with stabilizer group $G = \langle g_1,..., g_n \rangle$. Lemma \ref{lemma: stab pauli orthog} and Corollary \ref{cor: basis from stab} (p.\pageref{cor: basis from stab}) tell us that acting on $\ket{\psi}$ with a Pauli $P$ either fixes it, or takes it to an orthogonal state, in particular the one selected by negating all the $g_i$ with which $P$ anti-commutes. There are $2^n$ such states, forming a basis for $n$-qubit state space. We need to determine the image of $P\ket{\psi}$. 

To do so, we must determine with which of the generators $g_1,..., g_n$ the operator $P$ anti-commutes. By Lemma \ref{lemma: commuting symplectic} (p.\pageref{lemma: commuting symplectic}), these are the generators whose binary representations as $2n$ bit strings have symplectic inner product $1$ with the binary representation of $P$. If one tests whether $P$ anticommutes with each $g_j$, one can construct a binary vector $\ket{b} = (\langle P, g_1 \rangle, ..., \langle P, g_n \rangle)$, where the inner products are the symplectic inner product. Labeling basis states with the binary representations of the numbers $0, ..., 2^n-1$, we can tell what the image of each map will be. This condition can be rephrased in a way that is simpler to deal with. 

\begin{lemma}\label{lemma: stab near uniform}
Suppose $\hat{G} = \langle g_1,..., g_n \rangle$ is a stabilizer group, and $\ket{\psi}$ is the corresponding stabilizer state. Let ${G}$ be the $n\times 2n$ binary matrix with rows given by $g_1, ..., g_n$. Observe that $G$ specifies a linear subspace (and also a linear map $G: \ZZ_2^{2n}\rightarrow \ZZ_2^n$). Let $H$ be a generator matrix for the dual of the subspace specified by $G$ (that is, the kernel of the map $G$). Then for $a,b\in \{0,1\}^n$,  the image of $X^aZ^b\ket{\psi} = \ket{H(a,b)}$.
\end{lemma}
\begin{proof}
	For each generator $g_j = (s,t)$, where $s$ are the first $n$ bits, referring to the positions that $X$ acts, and $t$ the second $n$ bits, referring to the positions that $Z$ acts, let $h_j = (t,s)$. We claim that $\langle h_1, ..., h_n \rangle$ generates the kernel of the map $G$, and thus that the matrix $H$ with rows $h_i$ is a generator matrix for the dual of the space $G$.  To demonstrate this, we must show that the set $\{h_1,..., h_n\}$ is linearly independent, and that $h_j \cdot g_k = 0 \mod 2$ for every $j,k$ (that is, they are orthogonal). The first part follows as $\{g_1,..., g_n\}$ are linearly independent, and we obtain the $h_j$'s by reordering of columns of $g_j$. The second condition holds as $G$ is a stabilizer, and hence each of the symplectic inner products $\langle g_j, g_k \rangle = 0$. With $g_j = (s_j, t_j)$, for each $j,k$ we have
	\be
		0 &=&\langle g_j, g_k \rangle \\
		&=& s_j \cdot t_k + t_j \cdot s_k \\
		&=& g_j \cdot h_k
	\ee
which establishes the claim. Since the $H$ constructed defines a linear map with kernel $\hat{G}$, elements of distinct cosets of $\hat{G}$ have distinct images, and we can identify $X^aZ^b\ket{\psi} = \ket{H(a,b)}$.
\end{proof}

To prove Theorem \ref{thm: stab condition} (p.\pageref{thm: stab condition}), we need only use Lemma \ref{lemma: stab near uniform} and observe that every stabilizer can be taken as the dual of some other stabilizer. 

Thus, we have established that for in order for the map $\EE$ to be $\epsilon$-randomizing, the underlying set $\ESS$ must obey the condition that for every stabilizer generator matrix $G$, the distribution $G(\ESS)$ (see Theorem \ref{thm: stab condition}) must be at most $\epsilon$-away from uniform. Unfortunately, this does not mean that \emph{every} matrix obeys this condition, as there are many linear subspaces of $\ZZ_2^{2n}$ that do not correspond to valid stabilizer groups. Nevertheless, this last condition implies all our earlier ones, as the class of stabilizer states includes the Pauli eigenvectors, the cat states, and linear subspace states.

We hope that this more general correspondence may be used in proving a lower bound. Possible avenues of attack could be to come up with an interpretation for other $n\times 2n$ binary matrices in terms of states, and apply a similar technique. Alternatively, perhaps one can argue based on some kind of permutation symmetry of the small bias spaces $\ESS$.

\clearpage
\thispagestyle{empty}
\cleardoublepage

\chapter{Conclusion and Future Research}
\label{ch:conclusion}
\markright{Conclusion and Open Problems}

\section{Conclusion}

We have studied approximately randomizing maps in the trace norm. In the area of upper bounds, we have demonstrated the best known upper bounds on $\epsilon$-randomizing maps, and also that most randomly selected sets of operators of this size are also $\epsilon$-randomizing.

We have also established a lower bound that is dependent on $\epsilon$, but is not tight, and developed some techniques which may be useful in further work.

\section{Directions for Future Research and Open Problems}

The most obvious open problem is that of establishing a tight lower bound for $\epsilon$-randomizing maps in the trace norm. We are continuing work in this direction. Further exploration along the lines of approximate designs \cite{D06} may be a fruitful approach. One can also rephrase these problems in terms of noisy channels. For example, if one models a $d$-dimensional noisy channel using Kraus operators with the uniform distribution, what is the smallest number of Kraus operators that can prevent a transmission rate greater than $f(\epsilon)$ for some function $f$. We have not examined this approach at all, and it is possible that a consideration of these problems in this alternative framework will yield some interesting results.

Also open are the questions of developing explicit schemes in the $\infty$-norm, and of coming up with lower bounds valid in the $\infty$-norm. Kerenidis and Nagaj \cite{KN05} demonstrated that the explicit schemes of Ambainis and Smith \cite{AS04} based on small-biased spaces are not valid in the $\infty$-norm, so the results of Hayden \etal \cite{HLSW04} remain the best known.

\appendix


\clearpage
\thispagestyle{empty}
\cleardoublepage

\chapter{Mathematical Background}
\label{ch:background}
\markright{Mathematical Background}




This appendix presents some of the mathematical results and notation used elsewhere in this thesis. We assume a basic knowledge of linear algebra and the formalisms of quantum computation. For an introduction to the language and notation of quantum computation \cite{NC00} is an excellent resource. 
%

\section{Pauli Group and Pauli Eigenvectors}

\subsection{Pauli Group}\label{sec: pauli group}
The Pauli Group is of fundamental importance in Quantum Computation and possesses several  interesting qualities which we make use of throughout this work.

\begin{definition}
Let $\identity$, $X$, $Z$ and $Y$ represent the following matrices respectively:
$\left[ \begin{array}{cc} 1 & 0 \\ 0 & 1\end{array}\right]$, $\left[ \begin{array}{cc} 0 & 1 \\ 1 & 0\end{array}\right]$, $\left[ \begin{array}{cc} 1 & 0 \\ 0 & -1\end{array}\right]$, $\left[ \begin{array}{cc} 0 & -i \\ i & 0\end{array}\right]$. Define $\PP_1 = \{\pm\identity, \pm i \identity, \pm X, \pm i X, \pm Y, \pm i Y, \pm Z, \pm i Z\}$. We call $\PP_1$ the Pauli Group.
\end{definition}

Observe that $\PP_1$ is a subgroup of $\mathcal{U}(2)$. Each element other than the identity has order 2, and the following identities hold:
\be
XZ &=& -ZX\\
Y &=& iXZ \\
\tr(P) &=& 0 \qquad \text{for $P = X,Y,Z$}\\
P^2 &=& \identity \qquad \forall P \in \PP_1.
\ee
Thus, ignoring phases, each non-identity Pauli anti-commutes with each of the other non-identity Paulis. The Pauli matrices are all unitary. Observe further that $\PP_1$ forms a basis for $\LL(2)$ that is orthonormal with the inner product $\langle A , B \rangle = \frac{1}{2} A^\dagger B$. The Pauli operators correspond to rotations of $180^o$ about the three axes of the Bloch sphere.

Next we define a generalization of the Pauli operators to $n$-qubit systems. 
\begin{definition}\label{def: n-qubit paulis}
Let $\PP_n$ be the set of all $n$-wise tensor products of Pauli matrices. The elements of $\PP_n$ can thus be written as strings in $\{X,Y,Z,\identity\}^{n}$, again up to a possible phase of $\pm1, \pm i$. 

We let $X_j\in \PP_n$ be the operator  consisting of an $X$ acting on the $j$th qubit, and $\identity$ on each other qubit. That is,
\be
	X_j &=& \identity \otimes ... \otimes \identity \otimes X \otimes \identity \otimes ... \otimes \identity
\ee
and define $Y_j$ and $Z_j$ similarly. Then $\PP_n$ can be found as arbitrary products (including the empty product) of $\{X_j, Y_j, Z_j\}_{j=1,..., n}$ times the $\identity$.

We will frequently refer to elements of $\PP_n$ as Pauli operators or matrices. 
\end{definition}

Let $a$ be an $n$-bit string. Then for an operator $P$, we will say $P^0 = \identity$ and $P^1 = P$ and define
\be
	P^a &=& \bigotimes_{j=1}^n P^{a_j}.
\ee
Note that for any $P \in \PP_n$, we have $P = \alpha X^a Z^b$ for some strings $a,b$ and $\alpha \in \{1,-1,i,-i\}$. We will often ignore the initial phase factor (which is reasonable, as we will most often be conjugating with $P$), and write $P = X^aZ^b$.

For two non-identity, single qubit Pauli operators $M_j, N_k$ with $M,N\in\{ X,Y,Z\}$, and $j,k \in \{1,...,n\}$, they either commute or anti-commute according to the following relation.
\bes
	M_j N_k &=& (-1)^{(1-\delta_{M,N})\delta_{j,k}} N_k M_j\label{eq: commutation relation}
\ees

Alternatively, one can phrase this in terms of the symplectic inner product.
\begin{definition}\label{def: symplectic}
Given two vectors $(a,b)$ and $(c,d)$ in $\ZZ_2^n$, we define their \emph{symplectic inner product} as 
\be
	\langle (a,b), (c,d) \rangle &=& a\cdot d + b \cdot c \mod 2
\ee
where $(\cdot)$ is the standard dot product.
\end{definition}
\begin{lemma}\label{lemma: commuting symplectic}
Two Paulis $X^aZ^b, X^cZ^d \in \PP_n$ anti-commute if and only if the symplectic inner product $\langle (a,b) , (c,d) \rangle = 1$.
\end{lemma}
\begin{proof}
	This follows from Equation (\ref{eq: commutation relation}). In particular, given the following map:
	\be
		\identity=X^0Z^0 &\mapsto&00\\
		X =X^1Z^0&\mapsto& 10\\
		Y=iX^1Z^1 &\mapsto &11\\
		Z=X^0Z^1 &\mapsto &01
	\ee
splitting each of the right-hand sides in two, we have that the symplectic inner product is 0 if and only if the left-hand sides commute. Then for two elements of $\PP_n$ represented by $(a,b)$ and $(c,d)$, they will commute if and only if they have anti-commuting elements in an even number of positions, that is, if 
\be
	\langle (a,b), (c,d)\rangle &=& 0.
\ee
\end{proof}

One can apply an element $P$ of $\PP_n$ to a system of $n$-qubits by applying the Pauli operator in the $j$th position of the string representing $P$ to the $j$th qubit of the system. For example,
\be
	XZXY\ket{1111} &=&
		(X\otimes Z \otimes X \otimes Y) (\ket{1}\otimes\ket{1}\otimes\ket{1}\otimes\ket{1})\\
		&=& (X\ket{1}) \otimes (Z\ket{1}) \otimes (X\ket{1}) \otimes (Y\ket{1})\\
		&=& \ket{0} \otimes (-\ket{1}) \otimes \ket{0} \otimes (-i \ket{0})\\
		&=& i\ket{0100}.
\ee

\begin{lemma}\label{lemma: paulis orthogonal basis}
The elements of $\PP_n$ form an orthogonal basis for the space of linear operators on $2^n$ dimensions, $\mathcal{L}(2^n)$ under the inner product $\left<P,Q\right> = \tr(P^\dagger Q)$.
\end{lemma}
\begin{proof}
Let $A$ and $B$ be two elements of $\PP_n$. Then
\be
	\tr(A^\dagger B) &=& \tr\left( \bigotimes_{j=1}^n A_j^\dagger B_j\right) \\
		&=& \prod_{j=1}^n \tr(A_j^\dagger B_j) 
\ee
as tensor products commute with trace. Then as $\PP_1$ is an orthogonal set with each element having squared norm $2$, we have 
\be
	\tr(A^\dagger B) &=&  \begin{cases}
					2^n & \text{if $A_j = B_j$ for each $j$} \\
					0 & \text{otherwise}
					\end{cases}.
\ee
Thus $\PP_n$ is orthonormal with normalization $\frac{1}{2^{n/2}}$. 
\end{proof}

Note that the norm induced by this inner product is in fact the $2$-norm (Section \ref{sec: 2-norm}). As $\PP_n$ is a basis for $\mathcal{L}(2^n)$, for any operator $\rho$ we can write $\rho$ as a sum of elements of  $\PP_n$. In particular, we have
\be
	\rho &=& \sum_{a,b\in \ZZ_2^n} \frac{1}{2^n}\tr(Z^bX^a\rho)X^aZ^b.
\ee

\subsection{Pauli Eigenvectors}\label{sec: pauli eigenvectors}
\begin{definition} We refer to the eigenvectors of $\PP_n$ as Pauli eigenvectors. For $\PP_1$, they are $\ket{0}$, $\ket{1}$ (corresponding to the $Z$ operator), $\ket{+} = \frac{1}{\sqrt{2}}(\ket{0}+\ket{1})$, $\ket{-} = \frac{1}{\sqrt{2}}(\ket{0}-\ket{1})$ (corresponding to $X$), $\ket{+i} = \frac{1}{\sqrt{2}}(\ket{0}+i\ket{1})$, $\ket{-i} = \frac{1}{\sqrt{2}}(\ket{0}-i\ket{1})$ (corresponding to $Y$). Each of these may be written in density operator form as 
\be
\frac{\identity + (-1)^w V}{2}
\ee
where $w = 0,1$ and $V\in\{X,Y,Z\}$.
The eigenvectors of elements of $\PP_n$  are simply the tensor products of the eigenvectors of the operators in the tensor product.
For example, the eigenvectors of $XX \in \PP_2$ are $\{\ket{+}\ket{+}, \ket{+}\ket{-}, \ket{-}\ket{+}, \ket{-}\ket{-}\}$. We will sometimes refer to the density matrix representations of these states as Pauli eigenvectors, and in that case, when $\ket{\psi}$ is an eigenvector of $P$, we will have 
\be
	P\proj{\psi}P^\dagger &=& \proj{\psi}.
\ee
\end{definition}

Note that each non-identity element of $\PP_n$ has exactly two eigenvalues, $\pm1$, and can be written as the difference between the projectors onto each of its eigenspaces. In the case of $\PP_1$, this means that each non-identity operator is the difference between the density matrix representations of its eigenvectors.

\begin{lemma}
For each operator $P\in \PP_n$, the eigenvectors of $P$ form an orthogonal basis for the space $\CC^{2^n}$. 
\end{lemma}
\begin{proof}
This is obvious, as each $P$ is unitary, and hence normal. Thus unitarily diagonalizing $P$ yields its eigenvectors, which must span the state space.
\end{proof}

\subsection{Heisenberg-Weyl Operators}\label{sec: Heisenberg Weyl}
The Pauli operators have a natural generalization from 2 dimensions to $d$ dimensions. In a $d$ dimensional space, consider the operators
\be
	X\ket{c} &\mapsto& \ket{c + 1 \mod d} \\
	Z\ket{c} &\mapsto& e^{c(2\pi i/d)}\ket{c}.
\ee
Many of the same results hold for the Heisenberg-Weyl operators as for the Pauli operators, and often constructions on qubits using the Paulis can be generalized to arbitrary dimension using the Heisenberg-Weyl operators.

\section{Norms}\label{app: norms}

In this thesis, we make use of three matrix norms: the trace (or 1-) norm, the frobenius (or 2-) norm, and the operator (or infinity-) norm.

\begin{definition}\label{defn:norm}
Given a vector space $V$ over $\CC$, the function $\norm{\cdot}:V\rightarrow \reals^+\cup\{0\}$ is called a norm if and only if the following conditions are satisfied.
\begin{enumerate}
	\item $\norm{v} \ge 0$ $\forall v\in V$ and $\norm{v} = 0 \Leftrightarrow v = 0$.
	\item For every $v\in V$ and $c\in \CC$, $\norm{cv} = |c|\norm{v}$.
	\item For every $u,v\in V$, $\norm{u+v} \le \norm{u} + \norm{v}$
\end{enumerate}
\end{definition}

\subsection{2-Norm}\label{sec: 2-norm}

The 2-norm of a vector is the usual Euclidean distance measure. It is easy to compute in general, and can be easily generalized to matrices, and related to the trace norm by inequalities. We denote the vector norm simply by $|\cdot|$, rather than the $\norm{\cdot}$ used for matrices.

\begin{definition}\label{def: 2-norm}
	For a vector $v$ in a Hilbert space $\HH$, its \emph{$2$-norm}, $|v|_2$ is defined as
	\be
		|v|_2 &=& \sqrt{\langle v , v \rangle}
	\ee
	where $\langle \cdot, \cdot \rangle$ is the inner product in $\HH$.
	For our purposes, we will consider only finite dimensional spaces isomorphic to $\CC^d$ as a Hilbert space (with the usual inner product). In this case, if $v$ is written as a complex column vector $v = (v_1,..., v_d)^T$, then we have the following equivalent expressions of $|v|_2$.
	\be
		|v|_2 &=& \left(\sum_{j=1}^d |v_j|^2 \right)^{1/2} \\
		&=& \sqrt{v^\dagger v}.
	\ee
	As this is the only vector norm we consider, we may omit the subscript.
\end{definition}

To define the $2$-norm on matrices, also known as the Hilbert-Schmidt or the Frobenius norm, we observe that an $s\times t$ matrix $M$ can be written as a vector of length $st$ simply by writing the columns of $M$ on top of each other rather than side-by-side, and that we can then use the vector definition. That is,
\bes
	\fnorm{M} &=& \left(\sum_{j = 1... s \atop k = 1... t} |m_{j,k}|^2\right)^{1/2}.\label{eqn: fnorm def}
\ees
There is a convenient alternative formulation for the 2-norm of a matrix given by the following lemma.
\begin{lemma}
For a $s\times t$ matrix $M$,
\be		
	\fnorm{M} &=& \sqrt{\tr(M^\dagger M)}.
\ee
\end{lemma}
\begin{proof}
	Let $r_1,..., r_t$ be the columns of $M$. We have 
	\be
		\tr(M^\dagger M) &=& \sum_{j=1}^t (M^\dagger M)_{j,j} \\
		&=& \sum_{j=1}^t r_j^\dagger  r_j\\
		&=& \sum_{j=1}^t \sum_{k=1}^s M_{k,j}^*M_{k,j} \\
		&=& \sum_{j=1}^t \sum_{k=1}^s |M_{k,j}|^2
	\ee
	and taking square roots on both sides completes the proof.
\end{proof}
\begin{corollary}
One can also diagonalize $M^\dagger M$, and then observe that the diagonal elements are the squares of the singular values of $M$, and thus if $\{\lambda_i\}$ are the singular values of $M$ (with multiplicity), 
\be
	\fnorm{M}^2 = \sum_{\lambda_i} |\lambda_i|^2.
\ee
\end{corollary}
We also show the following simple lemma about the outer product of a vector with itself.
\begin{lemma}
	For a finite-dimensional vector $v$, the 2-norm of its outer product with itself, $\fnorm{vv^\dagger}$ obeys
	\be
		\fnorm{vv^\dagger} &=& |v|_2^2.
	\ee
\end{lemma}
\begin{proof}
We have
\be
	\fnorm{vv^\dagger}^2 &=& \tr(vv^\dagger (vv^\dagger)^\dagger) \\
	&=& \tr(vv^\dagger vv^\dagger) \\
	&=& |v^\dagger v|^2 \\
	&=& |v|^4
\ee 
and taking square roots completes the proof.
\end{proof}
As quantum states are unit vectors, we also have the following corollary.
\begin{corollary}
	For a pure quantum state $\ket{\psi}$, we have
	\be
		|\ket{\psi}|_2 &=& \fnorm{\ketbra{\psi}{\psi}}.
	\ee
\end{corollary}

\subsection{Infinity Norm}
The $\infty$-norm is also known as the operator norm, or largest singular value norm. Hayden~\etal~\cite{HLSW04} worked in the infinity norm and we include it here for completeness. 
\begin{definition}\label{def: infnorm}
	For a linear operator $M:\mathcal{H}_1 \rightarrow \mathcal{H}_2$, 
	\be
		\infnorm{M} &=& \sup_{v \in \mathcal{H}_1\atop v \neq 0} \frac{\left| M v \right|}{|v|} 
	\ee
	When $M$ is a square matrix, this is equivalent to the absolute value of the largest eigenvalue of $M$ (otherwise, the largest singular value). Letting $c$ be the largest eigenvalue, and $v$ a unit eigenvector with eigenvalue $c$, this last condition allows us to write
	\be
		\infnorm{M} &=& |cv| \\
		&=& |c(v^\dagger v)| \\
		&=& |v^\dagger (cv) | \\
		&=& |v^\dagger M v|
	\ee
	and thus when the $v$ are restricted to unit vectors:
	\begin{equation}
		\infnorm{M} = \sup_{|v|=1} |v^\dagger M v| \label{eqn: alt infnorm}.
	\end{equation}
\end{definition}

\subsection{Trace Norm}\label{sec: trnorm}
The trace norm (also called the 1-norm) is a matrix norm. It is the norm that we are most interested in throughout this thesis. As it is often difficult to calculate directly, we need some relations between it and the other norms we have discussed. Rather than the largest singular value, the trace norm tells us the sum of the singular values of a matrix. It is also a more practically useful norm, for reasons that will be explained in this section.

Before defining the trace norm, we need the notion of the square root and absolute value of a matrix.
\begin{definition}\label{def: trnorm}
	For a diagonal matrix $D$, with non-negative, real diagonal elements $d_{j,j}$, define $\sqrt{D}$ to be the matrix with diagonal elements $\sqrt{d_{j,j}}$. For $M$ a Hermitian, positive definite matrix, we can write $M = P D P^\dagger$ for some orthogonal $P$ and diagonal $D$. Further, the elements of $D$ are all positive. Then we define 
	\be
	\sqrt{M} &=& P \sqrt{D} P^\dagger.
	\ee
In other words, $\sqrt{M}$ is the matrix that has the same eigenbasis as $M$, but with eigenvalues that are the square roots of those of $M$.	
\end{definition}
\begin{definition}
	The absolute value of a matrix M is 
	\be
		|M| &=& \sqrt{M^\dagger M}.
	\ee
	Note that $M^\dagger M$ is always Hermitian and positive definite, so this is well defined. For a square matrix $M$, $|M|$ is the matrix with the same eigenvectors as $M$, but with eigenvalues that are the absolute values of those of $M$.
\end{definition}
We are now ready to define the trace norm of a matrix.
\begin{definition}
	For a matrix $M$, the trace norm $\trnorm{M}$ is defined as:
	\be
		\trnorm{M} &=& \tr\left(|M|\right)\\
			&=& \tr\left(\sqrt{M^\dagger M}\right).
	\ee
	and as the trace of a matrix is equal to the sum of its eigenvalues, we also have
	\be
		\trnorm{M} &=& \sum_{\lambda_j} |\lambda_j|
	\ee
	where the $\{\lambda_j\}$ are the eigenvalues of $M$.
\end{definition}

Our interest in the trace norm is motivated by the following theorem from \cite{NC00}.
\begin{theorem}
	Let $\rho$ and $\sigma$ be a density matrices and $\{E_m\}$ be a POVM. The measurement outcomes for $\rho$ and $\sigma$ are given by $p_m = \tr(E_m \rho)$ and $q_m = \tr(E_m \sigma)$ respectively. Then 
	\be
		\trnorm{\rho-\sigma} &=& \max_{\{E_m\}} \left(\sum_{j=1}^m |p_j - q_j|\right).
	\ee
\end{theorem}	
This theorem shows us that the trace distance between two density operators corresponds in a very real sense to their distinguishability. In particular, if two density operators are close in the trace norm, then the trace norm provides an upper bound on the difference between the outcome distributions. In the context of our work, we want to come up with operations on density matrices that render them indistinguishable from the totally mixed state (which has measurement outcomes corresponding to the uniform distribution) except with a small probability.

One disadvantage of the trace norm is that it can be hard to compute. As such, we sometimes wish to relate it to the 2-norm and the $\infty$-norm. The following lemma displays the relationships between these norms.

\begin{lemma}\label{lemma: norm relations}
	For a $d \times d$ matrix $M$, we have
	\begin{eqnarray}
		\fnorm{M} &\le& \trnorm{M} \label{eqn: 2<tr}\\
		\infnorm{M} &\le&\fnorm{M} \label{eqn: inf<2}\\
		\trnorm{M} &\le& \sqrt{d}\fnorm{M}\label{eqn: tr<2}\\
		\trnorm{M} &\le& d\infnorm{M}\label{eqn: tr<inf}
	\end{eqnarray}
\end{lemma}
\begin{proof}
	Let $\{\lambda_i \st i = 1 ... d\}$ be the singular values of $M$.\\
   (\ref{eqn: 2<tr}) 
	The claim corresponds to the statement 
	\be
		\sum_{\lambda_i} |\lambda_i|^2 &\le& \left(\sum_{\lambda_i} |\lambda_i|\right)^{2}.
	\ee
	This is obvious, as the expansion of the right hand side includes all the terms on the left, and all the other terms are non-negative.\\ \\
   (\ref{eqn: inf<2}) 
   	This is equivalent to 
	\be
		\max_i|\lambda_i|^2 &\le& \sum_{i} |\lambda_i|^2
	\ee 
	which is clearly true, as the left hand side is one of the terms on the right.\\ \\
   (\ref{eqn: tr<2}) 
   	Let $J$ be the $d$-dimensional vector with all entries equal to $1$. Let $\{\lambda_i \st i = 1... d\}$ be the singular values of $|M| = \sqrt{M^\dagger M}$, and $\lambda$ be the $d$-dimensional vector with $i$th element $\lambda_i$. By the Cauchy-Schwartz inequality, we have:
	\be
	|\langle \lambda , J \rangle |^2 &\le& {|\lambda|^2|}{|J|^2}
	\ee
  	Expanding, we obtain:
	\be
	\left(\sum_{j=1}^d \lambda_i\right)^2 &\le& \left(\sum_{j=1}^d \lambda_i^2\right) d.
	\ee
	As $\lambda_i = |\lambda_i|$, taking square roots completes the proof. \\ \\
   (\ref{eqn: tr<inf})  
   	As $\infnorm{M}$ is the largest singular value of $|M|$, and $\trnorm{M}$ is the sum of singular values, the result is clear.\\
\end{proof}

The following technical lemma relating the $2$-norm on vectors to the trace norm is used in the construction of $\epsilon$-nets in section \ref{sub: enet}.
\begin{lemma}\label{lemma: enet lemma}
	For two pure states ${\varphi}$ and ${\psi}$, 
	\be
		\fnorm{\ket{\varphi} - \ket{\psi}}^2  &\ge& \left(\frac{1}{2}\trnorm{{\varphi} -{\psi}}\right)^2.
	\ee
\end{lemma}
\begin{proof}
For two pure states $\psi$ and $\varphi$, we have the following identity (see \cite{NC00}, Section 9.2.3)
\bes
	\trnorm{\varphi-\psi} = 2\sqrt{1-\left|\braket{\varphi}{\psi}\right|^2}.\label{eqn: tr-fid}
\ees
Also, for two pure states $\psi$ and $\varphi$, 
	\bes
		\fnorm{\ket{\varphi} - \ket{\psi}}^2  &=& \braket{\varphi}{\varphi} - \braket{\varphi}{\psi} - \braket{\psi}{\varphi} + \braket{\psi}{\psi} \nonumber\\
		 &=&2 - 2\Re(\braket{\varphi}{\psi}). \label{eqn: f-dist}
	\ees
	Thus using (\ref{eqn: tr-fid}) and (\ref{eqn: f-dist}) we have
	\be
		\fnorm{\ket{\varphi} - \ket{\psi}}^2 &=& 2 - 2\Re(\braket{\varphi}{\psi})\\
		&\ge& 2 - 2|\braket{\varphi}{\psi}| \\
		&\ge& {1-|\braket{\varphi}{\psi}|^2}\\
		&=& \frac{1}{4}\trnorm{\varphi-\psi}^2
	\ee
	which establishes the lemma.
\end{proof}

One more fact that is useful to us is the unitary invariance of the trace norm.
\begin{definition}
	A matrix norm $\norm{\cdot}$ is said to be unitarily invariant if for every pair of unitary matrices $U,V$ (of the appropriate dimension),
	\be
		\norm{M} &=& \norm{UMV}.
	\ee
\end{definition}
\begin{lemma}\label{lemma: unitarily invariant}
	The trace norm is unitarily invariant.
\end{lemma}
\begin{proof}
The trace norm is the sum of the absolute values of the singular values of a matrix. One can thus use the singular value decomposition to see that the trace norm is not affected by unitary operations.
%
\end{proof}
%

\clearpage
\thispagestyle{empty}
\cleardoublepage

\chapter{The Stabilizer Formalism}
\label{chap: stabilizers}
\markright{The Stabilizer Formalism}

\section{Stabilizers}
The stabilizer formalism was introduced by Daniel Gottesman \cite{G97}. It provides an alternative view to that of kets and density matrices for looking at a large variety of states. It has been used widely in error correction.

\begin{definition}
An abelian subgroup of $\PP_n$ is called a \emph{stabilizer group} if it does not contain $-\identity$.
\end{definition}

\begin{definition}
Let $S<\PP_n$. Define $V_S = \{\psi \st M\psi = \psi \ \forall M \in S\}$. Then $V_S$ is the the set of $n$-qubit states that are invariant under the action of every element of $S$. We call $V_S$ the \emph{space stabilized by $S$}.
\end{definition}

Clearly $V_S= \emptyset$ for any group $S$ containing $-\identity$, which explains the restriction that stabilizer groups must not contain $-\identity$. It is simple to verify that any linear combination of elements of $V_S$ is also in $V_S$, and thus that $V_S$ is a linear subspace of the space of $n$-qubit quantum states. We also see that $V_S$ is the intersection of the eigenspaces of eigenvalue $1$ of each of the operators in $S$. 
 
For any subgroup  $S<\PP_n$, we can write a finite list of elements $g_1,..., g_k$ such that every element of $S$ can be written as a finite product of $g_1,..., g_k$. If $g_1, ..., g_k$ is a minimal such set, we say that $S$ is generated by $g_1,..., g_k$, and we write $S = \langle g_1,..., g_k \rangle$.

Ignoring phases, we can write each Pauli operator as a $2n$-bit string as in Definition \ref{def: n-qubit paulis}. Then multiplication of Paulis corresponds to addition modulo $2$ of their string representations and multiplication of their phases. Then a set of generators $g_1,..., g_k$ must have string representations that are linearly independent. 

Using this representation of generators, for a given stabilizer group $S = \langle g_1,..., g_k\rangle$ we can write them as the rows of a $k\times 2n$ binary matrix $M$. Then one can obtain all possible elements of $S$ by taking arbitary $\ZZ_2$ linear combinations of the rows of $M$. 

\begin{definition}
If $V_S$ is a $1$-dimensional space, then there is a single unit vector (up to phase) $\ket{\psi_S}\in V_S$ representing a unique quantum state. We call such states \emph{stabilizer states}.
\end{definition}

\begin{lemma}
If $S < \PP_n$ is a stabilizer group, then $V_S$ represents a stabilizer state if and only if $S$ is generated by $n$ generators.
\end{lemma}
\begin{proof}
This is a special case of Proposition 10.5 of \cite{NC00}.
\end{proof}

Stabilizer states are a large class of states that include many interesting families, including the Pauli eigenvectors, and Bell states. For a discussion, see \cite{NC00}, Chapter 10.

For our purposes, we are mostly interested in the action of Pauli operators on stabilizer states. In particular, we have the following lemmas.

\begin{lemma}\label{lemma: one generator}
Given a stabilizer group $G$, and an arbitrary Pauli operator $P$, either $P$ commutes with every element of $G$, or there is a set of generators $g_1,..., g_n$ of $G$ such that $P$ anti-commutes with $g_1$, and commutes with $g_2,..., g_n$.
\end{lemma}
\begin{proof}
	Choose an arbitrary set of generators for $G$, $h_1,..., h_n$. If $P$ commutes with every $h_i$, $i=1,...,n$, then $P$ commutes with every element of $G$. Otherwise, $P$ anti-commutes with at least $1$ generator, without loss of generality, $h_1$. Let $g_1=h_1$ and For each $i=2,...,n$, let
	 \be
	 	g_i &=& \begin{cases}
				h_i & \text{if $g_i$ commutes with $P$}\\
				h_1h_i & \text{if $g_i$ anti-commutes with $P$}.
				\end{cases} 
	\ee
Clearly, $P$ will commute with each $g_i$ for $i = 2,...,n$, and the lemma holds.
\end{proof}

\begin{lemma}\label{lemma: stab pauli orthog}
	Let $\ket{\psi}$ be an $n$-qubit stabilizer state with $\stab(\ket{\psi}) = G = \langle g_1, ..., g_n \rangle$. Then for any Pauli $P\in \PP_n$, we have
	\be
		\bra{\psi}P\ket{\psi} &=& \begin{cases}
							1 & \text{if $P \in G$}\\
							0 & \text{otherwise}
							\end{cases}.
	\ee
\end{lemma}
\begin{proof}
The first part is easy to see; as $P\in G$, we have $P\ket{\psi} = \ket{\psi}$, and the inner product is clearly 1. For the second part, first observe that if $P \notin G$, then it must anticommute with at least one of the generators of $G$, including without loss of generality, $g_1$. Then we have
	\be
		\bra{\psi} P \ket{\psi} &=& \bra{\psi} g_1 P \ket{\psi} \\
		&=& -\bra{\psi} P g_1 \ket{\psi} \\
		&=& -\bra{\psi} P \ket{\psi}
	\ee 
	and thus $\bra{\psi}P\ket{\psi} =0$ as claimed.
\end{proof}


We observe that choosing a stabilizer group can in fact be used to pick a basis for the space of $n$-qubit states using the following corollary of  Lemma \ref{lemma: stab pauli orthog}.

\begin{corollary}\label{cor: basis from stab}
If $S = \langle g_1,..., g_n\rangle$, then if $T = \langle (-1)^{b_1}g_1, (-1)^{b_2}g_2,..., (-1)^{b_n}g_n \rangle$, $S'$ is a stabilizer group, and $\braket{\psi_S}{\psi_T} = 0$ unless $b_1 = ... = b_n = 0$.
\end{corollary}
\begin{proof}
	Clearly $-g_i$ is not in $S$ for any $i$, as otherwise $(-g_i)(g_i) = -\identity \in S$, which is a violation of the definition of a stabilizer group. Thus Lemma \ref{lemma: stab pauli orthog} implies the result.
\end{proof}

As there are $2^n$ such orthogonal states, this set forms a basis for pure states on $n$-qubits.
\clearpage
\thispagestyle{empty}
\cleardoublepage

\chapter{Small Bias Spaces and $k$-wise Independence}\label{chap: small bias}
\markright{Small Bias Spaces and $k$-wise Independence}

\section{Definitions}\label{sec: smallbias defns}
Small bias spaces are the foundation of the known explicit constructions for $\epsilon$-randomizing maps. They are closely related to the existence of almost $k$-wise independent spaces. These spaces have been well-studied (\cite{ABI85}, \cite{NN90}, \cite{AGHP92}), in computer science where they have been used in the derandomization of algorithms.

\begin{definition}[$k$-wise independence]
Let $\ESS$ be a sample space of $n$-bit binary strings. Let $X_i$ be the random variable corresponding to the $i$th bit of a string sampled from $\ESS$. We say that $\{X_1,...X_n\}$ is $k$-wise independent if for every set $|W|\subset [n], |W| = k$, the set $\{X_{W_1}, ..., X_{W_k}\}$ is independent.
\end{definition}

Alon, Babai and Itai \cite{ABI85} prove a tight lower bound on the size of the sample space $\ESS$.

\begin{theorem}
If the random variables $X_1, ..., X_n$ are $k$-wise independent and not almost constant (that is, they do not take a single value with probability 1), then $|\ESS| \in \Omega(n^{\lfloor k/2 \rfloor})$.
\end{theorem}

They also show that there exists a sample space of this size. This implies that the size of $\ESS$ is polynomial only for constant $k$. In light of this, Naor and Naor \cite{NN90} introduced the notion of \emph{almost $k$-wise independent} spaces. The following definitions are given by Alon, Goldreich, H{\aa}stad and Peralta \cite{AGHP92} to specify what exactly we mean by almost independence.

\begin{definition}\label{defn: eaway}
Let  $\ESS\subset \{0,1\}^n$ be a sample space, and let $X=X_1, ..., X_n$ be sampled uniformly from $\ESS$. 
\begin{itemize}
   \item $\ESS$ is $(\epsilon,k)$-independent (in the max norm) if for any $W \subset [n]$ such 
   	that $|W|=k$, and any $k$-bit string $\alpha$, 
	\begin{eqnarray*}
    		\Pr\left[X_{W_1}, ..., X_{W_k}= \alpha \right] &\le& \epsilon
	\end{eqnarray*}
  \item $\ESS$ is $\epsilon$-away  (in the trace norm) from $k$-wise independence, or alternatively, almost $k$-wise independent, if for any
  	$W \subset [n]$ such that $|W|=k$,
  	\begin{eqnarray*}
		\sum_{\alpha \in \{0,1\}^k} \left| \Pr\left[X_{W_1}, ... , X_{W_k} = \alpha\right] - 2^{-k}\right| 
		&\le& \epsilon
	\end{eqnarray*}
\end{itemize} 
\end{definition}

Clearly if $\ESS$ is $\epsilon$-away from $k$-wise independent, then it is $(\epsilon,k)$-independent, and if $\ESS$ is $(\epsilon,k)$-independent, then it is at most $2^k\epsilon$-away from $k$-wise independent.

In order to construct small spaces having these almost independence properties, Naor and Naor \cite{NN90} observe that the following are equivalent:
\begin{enumerate}
	\item $X_1, ..., X_n$ are uniform and independent
	\item $\forall W\subset [n]$, $\Pr\left[\bigoplus_{i\in W} X_i = 1\right]= \frac{1}{2}$
\end{enumerate}
For either of the above conditions to be satisfied, the sample space must have size $2^n$. Naor and Naor \cite{NN90} relax the second condition to instead require that instead the parities should be \emph{almost} equally even and odd. Vazirani \cite{V86} defined the \emph{bias} of a random variable as follows:
\begin{definition}\label{defn: bias}
	Let $X$ be a binary random variable. The \emph{bias} of $X$ is
	$$\left|\prob{X=0} - \prob{X = 1}\right|. $$
\end{definition}
A random variable is said to be \emph{unbiased} if it has bias 0.Thus the second condition above is that each of the random variables corresponding to the parities of subsets of $X_1, ..., X_n$ should be unbiased. 

\begin{definition}
	The bias of a sample space $\ESS$ of binary strings of length $n$ with respect to a string $\alpha$ of length $n$ is defined to be:
	\begin{eqnarray*}
		\bias_\alpha (\ESS) &=& \expct_{s\in\ESS}\left[(-1)^{\alpha \cdot  s}\right]
	\end{eqnarray*}
\end{definition}

Essentially, $\alpha$ is a witness to the randomness of $\ESS$. If the bias with respect to some $\alpha$ is large, then the space does not behave sufficiently randomly. This motivates the following definition, which captures the ability of a space to  act randomly, with respect to linear tests. That is, if every subset of the bits of $\ESS$ has the ``nearly equally distributed parity" property, then the space approximates in some sense the uniform distribution.

\begin{definition}\label{defn: ebias}
A sample space $\ESS$ is said to be $\epsilon$-biased with respect to linear tests of size at most $k$ if for every $\alpha$ such that $|\alpha|\le k$ ($\alpha \neq 0$),  
\begin{eqnarray*}
	\bias_\alpha(\ESS) &\le& \epsilon
\end{eqnarray*}
That is, for every non-empty subset $W \in [n]$ of bit positions, such that $|W| \le k$, the bias of the random variable $Y_W = \bigoplus_{i \in W} X_i$ where $X = X_1,X_2,...,X_n$ is sampled from $\ESS$ (that is, the parity of the bits of $X$ specified by $W$) satisfies
\begin{eqnarray*}
	\bias(Y_W) &\le& \epsilon
\end{eqnarray*}
When $k=n$, we abbreviate and say that $\ESS$ is $\epsilon$-biased.
\end{definition}

For clarity, we include the following toy example. Consider the set $S = \{00,11\}$. The bias of $S$ with respect to $01$ is:
\begin{eqnarray*}
	|2(1/2 (00\cdot01) + 1/2 (11\cdot 01)) -1| &=& 0
\end{eqnarray*}
however, with respect to $11$, it is:
\begin{eqnarray*}
	|2(1/2 (00\cdot11) + 1/2 (11\cdot 11)) -1| &=& 1.
\end{eqnarray*}
Thus we can see that $S$ is not $\delta$-biased for any $\delta < 1$. 

As previously observed, if $\epsilon = 0$,  the bits are completely independent, and that the size of the sample space must be $2^n$.  The following theorem of Vazirani \cite{V86} connects $\epsilon$-biased spaces to $k$-wise independence.

\begin{theorem}
	If $\ESS$ is $\epsilon$-biased with respect to linear tests of size at most $k$, then $\ESS$ is $((1-2^{-k})\epsilon, k)$-independent (max norm) and $(2^k-1)^{1/2}\epsilon$-away (in the trace norm) from $k$-wise independence.
\end{theorem}
\begin{proof}
	This proof, from \cite{AGHP92}, is by simple Fourier analysis and the Cauchy-Schwartz inequality. Without loss of generality, consider random variables $X_1, ..., X_k$ corresponding to the first $k$ bit positions of strings in $\ESS$. We denote by $\ESS_k$ their joint probability distribution. For each $\alpha \in \{0,1\}^k$, let 
\begin{eqnarray*}
	p_\alpha &=& \prob{X_1, ..., X_k = \alpha}
\end{eqnarray*}
Define the discrete Fourier transform of the sequence $p_\alpha$ by
\begin{eqnarray*}
	c_\beta &=& \sum_{\alpha \in \{0,1\}^k} (-1)^{\alpha\cdot\beta} p_\alpha
\end{eqnarray*}
We have $c_0 = 1$, and for $\beta \neq 0$, 
\begin{eqnarray*}
	|c_\beta| = \bias_\beta(\ESS_k)
\end{eqnarray*}
and thus by assumption $|c_\beta| \le \epsilon$. By standard Fourier analysis we have:
\begin{eqnarray*}
	p_\alpha &=& 2^{-k}\sum_{\beta} (-1)^{\alpha \cdot \beta} c_\beta
\end{eqnarray*}
(this is simply the result of applying the Fourier transform twice), and 
\begin{eqnarray*}
	\sum_\alpha p_\alpha^2 &=& 2^{-k}\sum_\beta c_\beta^2.
\end{eqnarray*}
Since $c_0 = 1$, we have 
\begin{eqnarray*}
	\left|p_\alpha - 2^{-k}\right| 
		& = & 2^{-k} \left|\sum_{\beta \neq 0} (-1)^{\alpha \cdot \beta} c_\beta\right| \\
		&\le& 2^{-k}\left|(2^k -1)\epsilon\right|\\
		& = & (1-2^{-k})\epsilon
\end{eqnarray*}
which shows the first part. For the second part, define 
\begin{eqnarray*}
	p_\alpha' &=& p_\alpha - 2^{-k} 
\end{eqnarray*} 
and let $c_\beta'$ be the Fourier transform of the $p'$ sequence. Then $c_0' = 0$ and $c_\beta' = c_\beta$ for $\beta \neq 0$ and by Cauchy-Schwartz inequality we have
\begin{eqnarray*}
	\sum_\alpha \left| p_\alpha - 2^{-k} \right| 
		&\le& 2^{k/2}\left( \sum_\alpha \left(p_\alpha - 2^{-k}\right)^2\right)^{1/2}\\
		& = & 2^{k/2}\left( 2^{-k}\sum_{\beta\neq 0} c_\beta^2 \right)^{1/2} \\
		&\le& \left(2^k-1\right)^{1/2}\epsilon
\end{eqnarray*}
which establishes the theorem.
\end{proof}

Conversely, if $\ESS$ is $\epsilon$-away from $k$-wise independence, it is generally true only that $\ESS$ is $\epsilon$ biased with respect to linear tests of size at most $k$.

\section{Constructions}\label{sec: small bias construction}
Alon, Goldreich, H{\aa}stad and Peralta \cite{AGHP92} give three constructions for $\epsilon$-biased spaces. We make use of their final construction (modified by their remark) in our explicit construction of $\epsilon$-PQCs. This construction is based on arithmetic in $GF(2^r)$ and uses $2r$ bits to generate a sample space $\ESS \subset \{0,1\}^{rs}$ with size $|\ESS| = 2^{2r}$. The proof of Proposition \ref{prop: small-bias} is an adaptation of the proof given in \cite{AGHP92}  for their third construction, modified to include the improvement of their remark.

\subsubsection{Construction}
Represent elements of $GF(2^r)$ as binary strings of length $r$. Let $v_1, ..., v_r$ be any basis for $GF(2^r)$ (for example the standard basis identifying $v_j = 0^{j-1}10^{r-j}$). An element of $z\in \ESS$ is specified by two strings of length $r$, say $x$ and $y$. The $ij$th bit (with $i\in\{0,...,s-1\}, j\in\{1,...,r\}$) of $z$ is specified by:
\begin{eqnarray*}
	z_{ij} &=& (v_j*x^i) \cdot y
\end{eqnarray*}
where $a\cdot b$ is the usual scalar product between vectors and $*$ is the multiplication in $GF(2^m)$ (hereafter omitted).

\begin{proposition} \label{prop: small-bias}
The sample space $\ESS$ is $\frac{s-1}{2^r}$-biased.
\end{proposition}

\begin{proof}
	Let $x$ and $y$ be elements of $GF(2^r)$ represented as binary strings of length $r$, and $z(x,y)		\in \ZZ_2^{rs}$ the element of $\ESS$ specified by $x$ and $y$. Then the $(ir+j)$th bit of $z(x,y)$ is given by
	\begin{eqnarray*}
		z(x,y)_{i,j} &=& \sum_{k=0}^{r-1} (v_j x^i)_k y_k.
	\end{eqnarray*} 
	We want to show that $z(x,y)$ is unbiased with respect to strings $\alpha\in \ZZ_2^{rs}$, so fix such 	an $\alpha$. Then we wish to consider $\langle z(x,y) , \alpha \rangle$. Expanding, we have
	\begin{eqnarray*}
		\langle z(x,y) , \alpha \rangle 
			&=& \sum_{i=0}^{s-1}\sum_{j=1}^{r} \alpha_{i,j} z(x,y)_{i,j} 
	\end{eqnarray*}		
	and then by linearity,
	\begin{eqnarray}
			&=& \sum_{i=0}^{s-1}\sum_{j=1}^{r} \alpha_{i,j} \sum_{k=1}^{r} \left(v_j x^i\right)_k y_k\nonumber\\
			&=& \sum_{k=1}^{r}\sum_{i=0}^{s-1} \sum_{j=1}^{r} \left(\alpha_{i,j} v_j x^i\right)_k y_k\nonumber\\
			&=& \sum_{k=1}^{r}\left(\sum_{i=0}^{s-1}\sum_{j=1}^{r} \alpha_{i,j} v_j x^i\right)_k y_k. \label{eqn: small-bias inner prod}
	\end{eqnarray}
	Observe that 
	\begin{eqnarray*}
		p_\alpha(x) = \sum_{i=0}^{s-1}\sum_{j=0}^{r-1} \alpha_{ij} v_j x^i
	\end{eqnarray*} 
	is a polynomial in $x$ over $GF(2^r)$ (since $\{v_1,...,v_r\}$ is a basis), and rewrite Eq. \ref{eqn: small-bias inner prod} as
	\begin{eqnarray*}
		\sum_{k=0}^{r-1}\left(\sum_{i=0}^{s-1}\sum_{j=0}^{r-1} \alpha_{ij} v_j x^i\right)_k y_k
			&=& \left\langle p_\alpha(x), y \right\rangle.
	\end{eqnarray*}
	Now we consider the result of fixing $x=x_0$ and ranging over all values of $y$. There are two cases to consider, the first is that that $p_\alpha(x_0)\neq 0$. In this case the scalar product is evenly distributed over $0$ and $1$ as $y$ is uniform. The second case is that  $p_\alpha(x_0)= 0$, in which case the inner product is $0$ for each value of $y$. However, as $p_\alpha$ has degree $s-1$, it has at most $s-1$ roots, thus 
	\begin{eqnarray*}
		\bias[\left\langle p_\alpha(x), y \right\rangle] &=&
			 \left|\prob{\left\langle p_\alpha(x), y \right\rangle=0} -
			 \prob{\left\langle p_\alpha(x), y \right\rangle=1}\right| \\
		  &\le& \frac{1}{2^r}\left|\left(s-1 + \frac{2^r-s-1}{2}\right) - \left(\frac{2^r-s-1}{2}\right)\right|\\
		  &=& \frac{s-1}{2^r}
	\end{eqnarray*}
\end{proof}

\section{Bounds on the Size of $\epsilon$-biased Spaces}

In proving lower bounds for the size of sets of operators that are $\epsilon$-randomizing, we try to demonstrate that these sets imply the existence of $\epsilon$-biased spaces. For these connections to be useful, we need to have lower bounds on the size of $\epsilon$-biased spaces. The following result from \cite{AGHP92} gives a lower bound that is also tight in the range $\epsilon = 2^{-\Theta(n)}$.

\begin{proposition}\label{thm: e-bias lower bound}
Let $\epsilon < 1/2$ , $n$ be a positive integer, and $\ESS$ be an $\epsilon$-biased space of $n$ binary random variables (i.e., a distribution of strings in $\{0,1\}^n$). Let $m(n,\epsilon)$ be the minimum possible size of $\ESS$. Then for every such $n$ and $\epsilon$, we have:
\begin{eqnarray*}
	m(n,\epsilon) &\ge& \Omega\left(\min\left\{ \frac{n}{\epsilon^2 \log(1/\epsilon)}, 2^n \right\}    \right)
		\qquad \text{and,} \\
	m(n,\epsilon) &\le&  \bigoh\left(\min\left\{ \frac{n}{\epsilon^2}, 2^n, \frac{n^2}{\epsilon^2(\log_2(n/\epsilon))^2} \right\}\right).
\end{eqnarray*}
\end{proposition}



\clearpage
\thispagestyle{empty}
\cleardoublepage

\markright{Bibliography}


\end{document}